\documentclass[preprint,12pt]{article}
\usepackage{amssymb}
\usepackage{graphicx}
\usepackage{multirow}
\usepackage{caption}
\usepackage{subfig}
\usepackage{mathtools}
\begin{document}
\title
{Strongly magnetized cold electron degenerate gas: Mass-radius relation of the magnetized white dwarf}
\author{Upasana Das, Banibrata Mukhopadhyay\\
Department of Physics, Indian Institute of
  Science, Bangalore 560012, India\\
upasana@physics.iisc.ernet.in, bm@physics.iisc.ernet.in}
\maketitle
\begin{abstract}

We consider a relativistic, degenerate electron gas at zero-temperature under the influence of a strong, uniform, static magnetic field, neglecting any form of interactions. Since the density of states for the electrons changes due to the presence of the magnetic field (which gives rise to Landau quantization), the corresponding equation of state also gets modified. In order to investigate the effect of very strong magnetic field, we focus only on systems in which a maximum of either one, two or three Landau level(s) is/are occupied. This is important since, if a very large number of Landau levels are filled, it implies a very low magnetic field strength which yields back Chandrasekhar's celebrated non-magnetic results. The maximum number of occupied Landau levels is fixed by the correct choice of two parameters, namely the magnetic field strength and the maximum Fermi energy of the system. We study the equations of state of these one-level, two-level and three-level systems and compare them by taking three different maximum Fermi energies. We also find the effect of the strong magnetic field on the mass-radius relation of the underlying star composed of the gas stated above. We obtain an exciting result that, it is possible to have an electron degenerate static star, namely magnetized white dwarfs, with a mass significantly greater than the Chandrasekhar limit in the range $2.3-2.6M_{\odot}$, provided it has an appropriate magnetic field strength and central density. In fact, recent observations of peculiar Type Ia supernovae - SN 2006gz, SN 2007if, SN 2009dc, SN 2003fg - seem to suggest super-Chandrasekhar-mass white dwarfs with masses up to $2.4-2.8M_{\odot}$, as their most likely progenitors. Interestingly our results seem to lie within the observational limits. 

\end{abstract}


\vspace{0.2cm}




\section{Introduction}

\indent \indent Neutron stars are known to have high magnetic fields as large as $10^{12}$ G or more on their surfaces. Several magnetic white dwarfs have also been discovered with surface fields from about $10^{5}$ G to $10^{9}$ G \cite{kemp}, \cite{putney}, \cite{schmidt}, \cite{reimers}, \cite{jordan} \cite{nord} and the physics of these objects have also been studied from a long time \cite{jordan92}, \cite{angel}, \cite{chanmugam}. It is likely that stronger fields exist in the centers of neutron stars or even white dwarfs, the limit to which is set by the scalar virial theorem \cite{lai}:
\begin{equation}
2T + W + 3\Pi + \mathcal{M} = 0, 
\end{equation}
where $T$ is the total kinetic (rotational) energy, $W$ the gravitational potential energy, $\Pi$ arises due to the internal energy and $\mathcal{M}$ the magnetic energy. Since $T$ and $\Pi$ are both positive, the maximum magnetic energy can be compared to, but can never exceed, the gravitational energy in an equilibrium configuration. For a star of mass $M$ and radius $R$ this gives $(4\pi R^{3}/3)(B_{max}^{2}/8\pi) \sim GM^{2}/R$, or $B_{max} \sim 2 \times 10^{8} (M/M_{\odot})(R/R_{\odot})^{-2}$ G. For white dwarfs this limit is $10^{12}$ G. Ostriker and Hartwick \cite{ost} had constructed models of magnetic white dwarfs with magnetic field strength $B \sim 10^{12}$ G at the center but with a much smaller field at the surface. Thus high interior magnetic fields in white dwarfs, although rarely observed in nature so far, are not completely implausible. 

It was proposed by Ginzburg \cite{ginzburg} and Woltjer \cite{woltjer} that the magnetic flux $\phi_{B} \sim 4\pi B R^{2}$ of a star is conserved during its evolution and subsequent collapse to form a remnant degenerate star (flux freezing phenomenon). Thus degenerate stars of small size and large magnetic fields are expected to be formed from parent stars which originally could have quite high magnetic fields of the order $\sim 10^{8}$ G \cite{spitzer}, \cite{shapiro}. Thus the study of such highly magnetized degenerate stars will help us understand the origin and evolution of stellar magnetic fields.

The mass-radius relation for (non-magnetic) white dwarfs was first determined by Chandrasekhar \cite{chandra}. He obtained a maximum mass for stable white dwarfs, known as the famous Chandrasekhar limit ($\sim 1.44M_{\odot}$), such that electron degeneracy pressure is just adequate to counteract gravitational collapse of the star. Suh et al. \cite{suh} obtained the mass-radius relation for white dwarfs with $B \sim 4.4 \times 10^{11-13}$ G. They, however, worked in the weak field limit (thus ignoring Landau quantization) and applied Euler-MacLaurin expansion to the equation of state of a fully degenerate electron gas in a strong magnetic field \cite{lai}, in order to recover the usual equation of state in absence of a magnetic field. They found that both the mass and radius of magnetic white dwarfs increase compared to non-magnetic white dwarfs, having the same central density.

In this paper, we consider a relativistic, degenerate electron gas at zero-temperature under the influence of a strong, uniform, static magnetic field. We neglect any form of interactions between the electrons. We study the effect of strong magnetic field on the equation of state of the degenerate matter and consequently obtain the mass-radius relation for a collapsed static star that might be composed of such matter. In order to highlight the effect of Landau quantization of electrons due to strong magnetic field, we restrict our systems to have at most one, two or three Landau level(s).

We hypothesize the possibility of existence of purely electron degenerate stars with extremely high magnetic fields $\sim 10^{15}-10^{17}$ G at the center, plausibly strongly magnetized white dwarfs. We also investigate the possibility of such stars having a mass greater than the Chandrasekhar limit, in the range $2.3-2.6M_{\odot}$. In a simple 
analytical framework existence of such stars has already been reported recently \cite{mpla} and its 
astrophysical implications based on numerical analysis was also discussed \cite{grf}. Interestingly recent observations of peculiar Type 1a supernovae - SN 2006gz, SN 2007if, SN 2009dc, SN 2003fg - seem to suggest super-Chandrasekhar-mass white dwarfs as their most likely progenitors \cite{nature}, \cite{scalzo}. These white dwarfs are believed to have masses up to $2.4-2.8M_{\odot}$. Proposed mechanism by which these white dwarfs exceed the Chandrasekhar limit is chiefly mass accretion from a binary companion accompanied by differential rotation \cite{kato}. This is fundamentally different from what we are proposing here. However, these observations are quite stimulating as not only do they support the existence of super-Chandrasekhar mass white dwarfs, but also our results seem to lie within the observational limits.

The paper is organized as follows. In the next section, we first recall how the equation of state of a cold electron degenerate gas gets modified due to the presence of a strong magnetic field and then state the numerical procedure followed to obtain results. Subsequently in \S3 we discuss the numerical results describing the nature of the equations of state and the mass-radius relations. In \S4 we elaborate on some of the key points of this work, for example, 
the timescale of magnetic field decay, comparison with Chandrasekhar's standard results, unstable branch of the 
mass-radius relations, the justifications for a constant magnetic field, the non-relativistic equation for 
hydrostatic equilibrium, neglecting Coulomb interactions and the anisotropy in pressure due to strong 
magnetic field. Finally, in \S5 we summarize our findings with conclusions. 

\section{Equation of state for a free electron gas in a strong magnetic field}
\subsection{Basic equations}
\indent \indent The energy states of a free electron in a uniform magnetic field are quantized into what is known as Landau orbitals, which define the motion of the electron in a plane perpendicular to the magnetic field. On solving the Schr\"{o}dinger equation in an external, uniform and static magnetic field directed along the $z$-axis, one obtains the following dispersion relation \cite{landau}
\begin{equation}
E_{\nu,\,p_{z}} = \nu \hbar \omega_{c} + \frac{p_{z}^{2}}{2m_{e}} ,
\label{1}
\end{equation}
where quantum number $\nu$ denotes the Landau level and is given by
\begin{equation}
\nu = j + \frac{1}{2} + \sigma , 
\end{equation}
when $j$ being the principal  quantum number of the Landau level ($j$ = 0, 1, 2,...), $\sigma = \pm \frac{1}{2}$, the spin of the electron, $m_{e}$ the rest mass of the electron, $\hbar$ the Planck's constant and $p_{z}$ the momentum of the electron along the $z$-axis which may be treated as continuous (the motion along the field is not quantized). The cyclotron energy is $\hbar \omega_{c} = \hbar (eB/m_{e}c)$, where $e$ is the charge of the electron, $c$ the speed of light and $B$ the magnetic field. 

\noindent Now, the electrons can become relativistic in either of the two cases:

(i) when the density is high enough such that the mean Fermi energy of an electron exceeds its rest-mass energy, 

(ii) when the cyclotron energy of the electron exceeds its rest-mass energy. 

\noindent We can define a critical magnetic field strength $B_{c}$ from the relation $\hbar \omega_{c} = m_{e}c^{2}$, which gives $B_{c} = m_{e}^{2}c^{3}/\hbar e = 4.414 \times 10^{13}$ G. Thus in order to study the effect of a strong magnetic field ($B \gtrsim B_{c}$) on the equation of state of a relativistic, degenerate electron gas we have to solve the relativistic Dirac equation. We mention here that for the present purpose, the magnetic field considered in the electron degenerate star originates due to the flux freezing phenomenon during the gravitational collapse of the parent star. The energy eigenstates in this case turn out to be \cite{lai}
\begin{equation}
E_{\nu,\,p_{z}} = [p_{z}^{2}c^{2} + m_{e}^{2}c^{4}(1 + \nu \frac{2B}{B_{c}})]^{1/2}.
\label{2}
\end{equation}
The main effect of the magnetic field is to modify the available density of states for the electrons. The number of states per unit volume in an interval $\Delta p_{z}$ for a given Landau level $\nu$ is $g_{\nu}(eB/h^{2}c)\Delta p_{z}$, where $g_{\nu}$ is the degeneracy that arises due to the Landau level splitting, such that, $g_{\nu} = 1$ for $\nu = 0$ and $g_{\nu} = 2$ for $\nu \geq 1$. Therefore the electron state density in the absence of magnetic field
\begin{equation}
\frac{2}{h^{3}}\int d^{3}p ,
\end{equation}
has to be replaced with 
\begin{equation}
\sum_{\nu} \frac{2eB}{h^{2}c}\,g_{\nu} \int dp_{z} 
\label{dos}
\end{equation}
in case of a non-zero magnetic field.

Now in order to calculate the electron number density $n_{e}$ at zero temperature, we have to evaluate the integral in equation (\ref{dos}) from $p_{z} = 0$ to $p_{F}(\nu)$, which is the Fermi momentum of the electron for the Landau level $\nu$, to obtain \cite{lai}
\begin{equation}
n_{e} = \sum_{\nu=0}^{\nu_{m}} \frac{2eB}{h^{2}c}\,g_{\nu}\,p_{F}(\nu).
\label{num}
\end{equation}
The Fermi energy $E_{F}$ of the electrons for the Landau level $\nu$ is given by
\begin{equation}
E_{F}^{2} = p_{F}(\nu)^{2}c^{2} + m_{e}^{2}c^{4}(1 + 2\nu \frac{B}{B_{c}}).
\label{fermi}
\end{equation}
The upper limit $\nu_{m}$ of the summation in equation (\ref{num}) is derived from the condition that $p_{F}(\nu)^{2} \geq 0$, which implies $E_{F}^{2} \geq m_{E}^{2}c^{4}(1 + 2\nu \frac{B}{B_{c}})$ and we obtain


\begin{equation}
\nu \leq \frac{\epsilon_{F}^{2} - 1}{2B_{D}} 
\end{equation}
or
\begin{equation}
\nu_{m} = \frac{\epsilon_{Fmax}^{2} - 1}{2B_{D}},
\label{max}
\end{equation}
where $\epsilon_{F} = E_{F}/m_{e}c^{2}$, is the dimensionless Fermi energy, $B_{D} = B/B_{c}$, the dimensionless magnetic field and $\epsilon_{Fmax} = E_{Fmax}/m_{e}c^{2}$, the dimensionless maximum Fermi energy of a system for a given $B_{D}$ and $\nu_{m}$. We note that $\nu_{m}$ is taken be the nearest lowest integer in equation (\ref{max}). For example, if $0 \leq \nu_{m} < 1$ for a particular value of $\epsilon_{Fmax}$ and $B_{D}$, then the upper limit is taken to be $\nu_{m}=0$.

If we define a dimensionless Fermi momentum $x_{F}(\nu) = p_{F}(\nu)/m_{e}c$, then equations (\ref{num}) and (\ref{fermi}) may be written as 
\begin{equation}
n_{e} = \frac{2B_{D}}{(2\pi)^{2} \lambda_{e}^{3}} \sum_{\nu=0}^{\nu_{m}} g_{\nu}x_{F}(\nu)
\label{ne}
\end{equation}
and
\begin{equation}
\epsilon_{F} = [x_{F}(\nu)^{2} + 1 + 2\nu B_{D}]^{1/2} 
\end{equation}
or
\begin{equation}
x_{F}(\nu) = [\epsilon_{F}^{2} - (1 + 2\nu B_{D})]^{1/2},
\end{equation}
where $\lambda_{e} = \hbar/m_{e}c$, is the Compton wavelength of the electron. The matter density $\rho$ can be written as
\begin{equation}
\rho = \mu_{e}m_{H}n_{e},
\label{mat}
\end{equation}
where $\mu_{e}$ is the mean molecular weight per electrons and $m_{H}$ the mass of hydrogen atom.

\noindent The electron energy density at zero temperature is
\begin{eqnarray}  
\varepsilon_{e} & = & \frac{2B_{D}}{(2\pi)^{2}\lambda_{e}^{3}}\sum_{\nu=0}^{\nu_{m}} g_{\nu} \int\limits_{0}^{x_{F}(\nu)} E_{\nu,\,p_{z}}d \left (\frac{p_{z}}{m_{e}c} \right) \nonumber \\ 
& = & \frac{2B_{D}}{(2\pi)^{2}\lambda_{e}^{3}}m_{e}c^{2}\sum_{\nu=0}^{\nu_{m}} g_{\nu} (1 + 2\nu B_{D}) \bold{\psi} \left (\frac{x_{F}(\nu)}{(1 + 2\nu B_{D})^{1/2}} \right), 
\label{endensity}
\end{eqnarray}
where
\begin{equation}
\bold{\psi}(z) = \int\limits_{0}^{z} (1 + y^{2})^{1/2}dy = \frac{1}{2}z \sqrt{1 + z^{2}} + \frac{1}{2}\ln(z + \sqrt{1 + z^{2}}).
\end{equation}
Then the pressure of an electron gas in a magnetic field is given by
\begin{eqnarray}
P_{e} & = & n_{e}^{2}\frac{d}{dn_{e}}\left (\frac{\varepsilon_{e}}{n_{e}} \right) = - \varepsilon_{e} + n_{e}E_{F} \nonumber \\
& = & \frac{2B_{D}}{(2\pi)^{2}\lambda_{e}^{3}}m_{e}c^{2}\sum_{\nu=0}^{\nu_{m}} g_{\nu} (1 + 2\nu B_{D}) \bold{\eta} \left (\frac{x_{F}(\nu)}{(1 + 2\nu B_{D})^
{1/2}} \right), 
\label{pressure}
\end{eqnarray}
where
\begin{equation}
\bold{\eta}(z) = \frac{1}{2}z \sqrt{1 + z^{2}} - \frac{1}{2}\ln(z + \sqrt{1 + z^{2}}).
\end{equation}

\subsection{Procedure}

\indent \indent Referring to the qualitative discussion by Lai and Shapiro \cite{lai}, let $\nu_{m}$ also denote the maximum number of Landau levels occupied by a cold gas of electrons in a magnetic field. In this case, from equation (\ref{max}), $\nu_{m}$ will be the nearest highest integer. Then if $\nu_{m} \gg 1$ then the Landau energy level spacing becomes a very small fraction of the Fermi energy and the discrete sum over $\nu$ can be replaced by an integral and we get back the non-magnetic results. From equation (\ref{max}) we see that if the magnetic field strength is high (for a fixed Fermi energy), i.e., $B_{D} \gg 1$, then $\nu_{m}$ is small and the electrons are restricted to the lower Landau levels only. It is in this case that the magnetic field plays an important role in influencing the equation of state for the relativistic degenerate gas. 

Since we are investigating the effects of high magnetic field in this work, we fix $\nu_{m}$, such that it can only take values 1, 2 or 3, which we call as one-level, two-level and three-level system respectively. To clarify further, by one-level system we mean where only the ground Landau level, $\nu = 0$, is occupied, two-level means where both the ground and the first ($\nu = 1$) Landau levels are occupied and three-level means where the ground, first and second ($\nu = 2$) Landau levels are occupied. Now from equation (\ref{max}) we see that once we fix $\nu_{m}$, we 
obtain a fixed $B_{D}$ on supplying a desired $E_{Fmax}$, which corresponds to the maximum possible 
density (in a star which corresponds to its central density) for that $\nu_m$. 
Hence, in our framework, the magnetic field is in accordance with the density of the system.
We choose $E_{Fmax} = 2\, m_{e}c^{2}$, $20\, m_{e}c^{2}$ and $200\, m_{e}c^{2}$ and for each we study the one-level, two-level and three-level systems, giving a total of 9 cases which are listed in Table 1. We mention here that for a given value of $E_{Fmax}$, the value of $B_{D}$ listed here corresponds to a lower limit. For example, when $E_{Fmax} = 20\,m_{e}c^{2}$, $B_{D}$ with a value of 199.5 just results in a one-level system but, if we choose any $B_{D} >$ 199.5 that would also lead to a one-level system.

For each of these cases we obtain the equation of state by simultaneously solving equations (\ref{ne}), (\ref{mat}) and (\ref{pressure}) numerically from $E_{F} =  m_{e}c^{2}$ to $E_{F} = E_{Fmax}$, when each value of $E_{F}$ gives one point in the $P_{e}-\rho$ plot. In this work we choose $\mu_{e} = 2$ throughout. Figure 1 shows the equations of state for the different cases given in Table 1, which will be discussed in \S3.

\begin{table}
\vskip0.2cm
{\centerline{\large Table 1}}
{\centerline{ Parameters for the equations of state in Figure 1. }}
{\centerline{}}
\begin{center}
\begin{tabular}{|c|c|c|c|}

\hline
$E_{Fmax}$ & $\rm{Maximum~Landau~level(s)}~\nu_{m}$ &  $B_{D}$ & $B$ in units of $10^{15}$ G\\
\hline

& 1 & 1.5 & 0.066 \\
 $2\,m_{e}c^{2}$ & 2 & 0.75 &  0.033 \\
& 3 & 0.5 &  0.022 \\ \hline
& 1 & 199.5 &  8.81 \\
 $20\,m_{e}c^{2}$ & 2 & 99.75 &  4.40 \\
& 3 & 66.5 & 2.94  \\ \hline
& 1 & 19999.5 & 882.78 \\
$200\,m_{e}c^{2}$ & 2 & 9999.75 & 441.38 \\
& 3 & 6666.5 &  294.26 \\ \hline

\end{tabular}

\end{center}

\end{table}

If we are to construct the model of a strongly magnetized star made out of electron degenerate matter, 
which is approximated to be spherical in presence of constant magnetic field, we require to solve the following 
differential equation which basically comes from the condition of hydrostatic equilibrium \cite{arc}
\begin{equation}
\frac{1}{r^{2}}\frac{d}{dr}\left(\frac{r^{2}}{\rho}\frac{dP}{dr} \right) = -4\pi G \rho,
\label{diff}
\end{equation}
where we consider $P = P_{e}$ throughout this work. See, however, the Appendix in order to understand 
the effect of deviation from spherical symmetry due to the anisotropic effects of magnetic field, as discussed
in \S 4.7.
Since the pressure cannot be expressed as an analytical function of density, unlike that of Chandrasekhar's work \cite{chandra}, we fit the equation of state with the following polytropic relation:
\begin{equation}
 P = K\rho^{\Gamma},
\label{poly}
\end{equation}
with different values of the adiabatic index $\Gamma$ in different density ranges ($K$ being a dimensional constant). Thus the actual equation of state is reconstructed using multiple polytropic equations of state. One such fit is shown in Figure 1(d) and the parameters $K$ and $\Gamma$ are stated in Table 2. The motivation behind doing such a fit is the following. First of all with this fitting we can determine the effect of magnetic field on the adiabatic index of the matter. More importantly, once we use an equation of state of the form (\ref{poly}), the problem essentially reduces to solving the Lane-Emden equation which arises in the non-magnetic case, except that in our case, $K$ and $\Gamma$ also carry information about the magnetic field in the system. 

We briefly recall here the Lane-Emden equation, since we would be referring to some of its solutions in the next section. We start with equations (\ref{diff}) and (\ref{poly}), and write $\Gamma = 1 + \frac{1}{n}$, where $n$ is the polytropic index. Next, two variable transformations are made as follows \cite{arc}: 

\begin{equation}
\rho = \rho_{c}\theta^{n},
\label{theta}
\end{equation}
where $\rho_{c}$ is the central density of the star and $\theta$ is a dimensionless variable
and
\begin{equation}
r = a\xi,
\label{xi}
\end{equation}
where $\xi$ is another dimensionless variable and $a$ is defined as
\begin{equation}
a = \left [\frac{(n+1)K\rho_{c}^{\frac{1-n}{n}}}{4\pi G} \right ]^{1/2},
\label{a}
\end{equation}
which has the dimension of length. Thus using equations (\ref{poly}), (\ref{theta}), (\ref{xi}) and (\ref{a}), equation (\ref{diff}) reduces to the famous Lane-Emden equation:
\begin{equation}
\frac{1}{\xi^{2}}\frac{d}{d\xi}\left(\xi^{2} \frac{d\theta}{d\xi} \right) = - \theta^{n},
\label{lane}
\end{equation}
which can be solved for a given $n$, subjected to the following two boundary conditions:
\begin{equation}
\theta(\xi = 0) = 1
\label{bc1}
\end{equation}
and
\begin{equation}
\left (\frac{d\theta}{d\xi} \right)_{\xi=0} = 0.
\label{bc2}
\end{equation}
If $n < 5$, then $\theta$ falls to zero for a finite value of $\xi$, called as $\xi_{1}$, which basically denotes the surface of the star where the pressure goes to zero (and then density becomes zero too in the present context). The physical radius of the star is then given by
\begin{equation}
R = a\xi_{1}.
\label{R}
\end{equation}
We note that the value of $n$ must be such that $n \geq -1$, so that $a$ is real in equation (\ref{a}) and $R \geq 0$ in equation (\ref{R}). Each value of central density $\rho_{c}$ corresponds to a particular value of radius $R$ and mass $M$ of a star. Substituting $a$ from equation (\ref{a}) in equation (\ref{R}) we find

\begin{equation}
R \propto \rho_{c}^{\frac{1-n}{2n}} = \rho_{c}^{\frac{\Gamma - 2}{2}}.
\label{rvsrho}
\end{equation} 

\noindent The mass of the spherical star (see, however, the Appendix discussing the equations for 
an oblate spheroid) is obtained by integrating the following equation:
\begin{equation}
\frac{dM}{dr} = 4\pi r^{2}\rho.
\label{mass}
\end{equation}
Hence,
\begin{equation}
M = 4\pi\int \limits_{0}^{R} r^{2}\rho \, dr = 4\pi a^{3} \rho_{c}\int \limits_{0}^{\xi_{1}} \xi^{2}\theta^{n}\, d\xi.
\label{M}
\end{equation}
Again substituting $a$ from equation (\ref{a}) in equation (\ref{M}) we find

\begin{equation}
M \propto \rho_{c}^{\frac{3-n}{2n}} = \rho_{c}^{\frac{3\Gamma -4}{2}},
\label{mvsrho}
\end{equation}
and then combining equations (\ref{rvsrho}) and (\ref{mvsrho}) we obtain the following mass-radius relation:

\begin{equation}
R \propto M^{\frac{1-n}{3-n}} = M^{\frac{\Gamma -2}{3\Gamma -4}}.
\label{rvsm}
\end{equation}

In the present work we do not make the transformations (\ref{theta}) and (\ref{xi}), but directly solve equation (\ref{diff}) (in different density regions corresponding to the particular values of $\Gamma$ or $n$) subjected to the same boundary conditions as in equations (\ref{bc1}) and (\ref{bc2}) which are written as

\begin{equation}
\rho(r = 0) = \rho_{c} 
\end{equation}
and
\begin{equation}
\left (\frac{d\rho}{dr} \right)_{r=0} = 0.
\end{equation}

\noindent Hence the results in equations (\ref{rvsrho}), (\ref{mvsrho}) and (\ref{rvsm}) still remain applicable to our case. A plot of $R$ as a function of $M$ gives the mass-radius relation for the magnetic, degenerate static star. 

Figure 2 shows the mass-radius relations for the one-level, two-level and three-level systems with $E_{Fmax} = 20\,m_{e}c^{2}$ (results for $E_{Fmax} = 2\, m_{e}c^{2}$ and $200\, m_{e}c^{2}$ also show the same trend). Figure 3 shows a comparison between the mass-radius relations for all the cases stated in Table 1. All these are discussed in detail in the next section.


\section{Numerical results}

\subsection{Equations of State}

\captionsetup[subfigure]{position=top}
\begin{figure}
  \begin{center}
    \begin{tabular}{ll}
 \subfloat[]{\includegraphics[scale=1.]{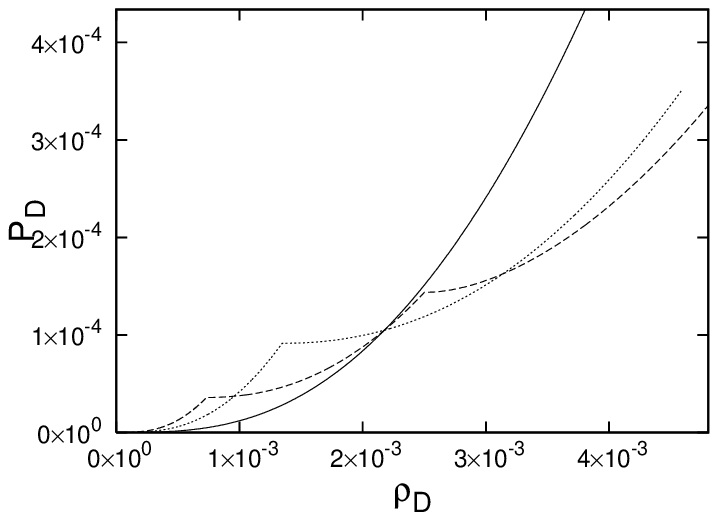}}&
   \subfloat[]{\includegraphics[scale=1.]{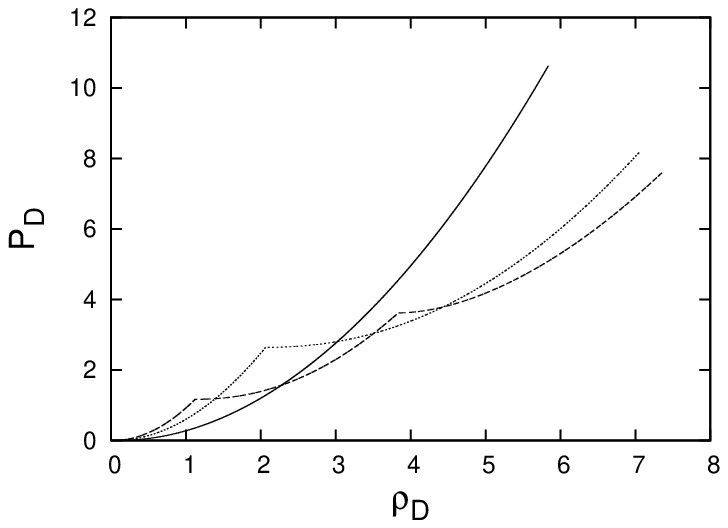}} \\ \\ \\
    \subfloat[]{\includegraphics[scale=1.]{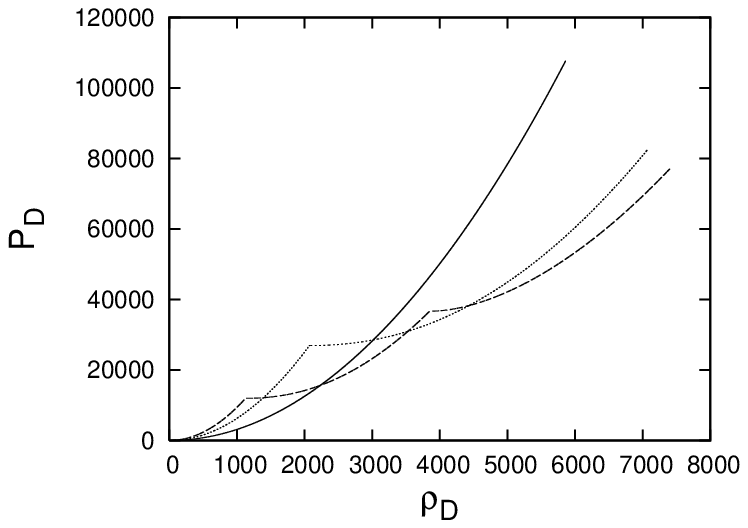}}&
 \subfloat[]{\includegraphics[scale=1.]{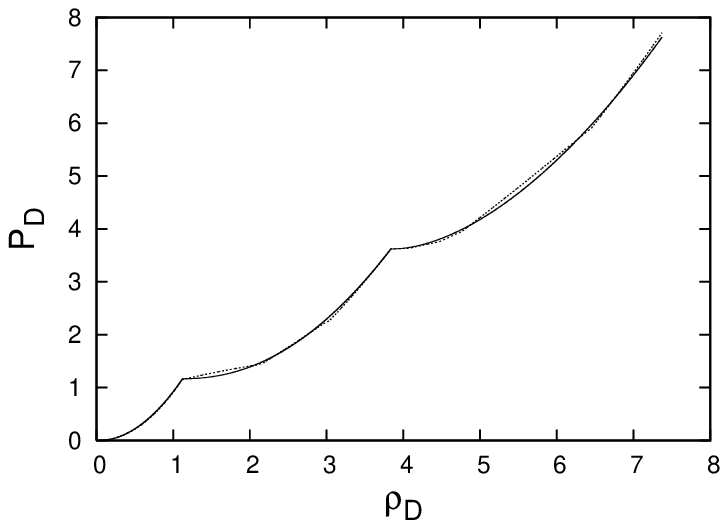}} \\
      \end{tabular}
    \caption{Equations of state in a strong magnetic field (given in Table 1) for (a) $E_{Fmax} = 2\,m_{e}c^{2}$, (b) $E_{Fmax} = 20\,m_{e}c^{2}$, (c) $E_{Fmax} = 200\,m_{e}c^{2}$. In all three cases the solid line, the dotted line and the dashed lines indicate one-level, two-level and three-level systems respectively. In (d) the solid line is same as the dashed line in (b), but fitted with the dotted line by analytical formalism (see text for details). Here $P_{D}$ is the pressure in units of $2.668 \times 10^{27}$ erg/cc and $\rho_{D}$ is the density in units of $2 \times 10^{9}$ gm/cc.}
    \label{eos}
  \end{center}
\end{figure}

\indent \indent Now we come to the discussions of the results obtained. We start with the equations of state shown in Figure 1. Let us consider the panel (b) which shows the cases with $E_{Fmax} = 20\, m_{e}c^{2}$. From Table 1 we see that the one-level system (the solid line) for $E_{Fmax} = 20\, m_{e}c^{2}$ corresponds to a magnetic field strength $B_{D} = 199.5$, and the two-level and three level systems (the dotted and dashed lines respectively) correspond to $B_{D} = 99.75$ and $ 66.5$ respectively.
We notice that the solid curve is free of any kink, the dotted curve has one kink and the dashed curve has two kinks. The kinks appear when there is a transition from a lower Landau level to the next and they demarcate regions of the equation of state where the pressure becomes briefly independent of density. Let us consider the two-level systems. The portion of the equation of state below the kink represents the ground Landau level and the one above the kink represents the first Landau level. As the Fermi energy of the electrons increases, more and more electrons occupy the ground Landau level and both the density and pressure of the system keep increasing. Once the ground level is completely filled, one observes that, on increasing the Fermi energy of the electrons the density increases but, the pressure remains fairly constant for a while, after which the pressure again starts increasing with density. It is as if, the increase in Fermi energy during the transition is being used by the system to move to a higher Landau level instead of increasing the pressure. This situation seems analogous to that of phase transition in matter (where the temperature remains constant with respect to the input heat energy during the change of phase). Similar features are also seen in Figures 1(a) and (c) for $E_{Fmax} = 2\, m_{e}c^{2}$ and $200\, m_{e}c^{2}$ respectively.

Thus looking at Figure 1, we observe that in the one-level systems (solid lines), all the electrons are in the ground Landau level and hence there is no kink. In the two-level systems (dotted lines), as the electrons start filling up the first Landau level, a kink develops in the equation of state. Finally in the three-level systems (dashed lines), there are two kinks $-$ the one at the lower density indicating transition to the first Landau level and another at the higher density indicating transition to the second Landau level. The value of $E_{Fmax}$ determines the maximum density of the system and hence the positions of the kinks shift accordingly in Figures 1(a), (b) and (c).

\subsection{Mass-radius relations}

\indent \indent Next we come to Figure 2. Here we show the mass-radius relations for the one-level, two-level and three level systems with $E_{Fmax} = 20\,m_{e}c^{2}$ (the explanation that follows also holds true for the cases with $E_{Fmax} = 2\,m_{e}c^{2} \textrm{ and } 200\,m_{e}c^{2}$). Each point in the mass-radius curve corresponds to a star with a particular value of central density $\rho_{c}$ which is supplied by us as a boundary condition ($R_D$ and $M_{D}$ are the dimensionless radius and mass of a star respectively as defined in Figure 2 caption).

\captionsetup[subfigure]{position=top}
\begin{figure}
  \begin{center}
    \begin{tabular}{ll}
 \subfloat[]{\includegraphics[scale=1.]{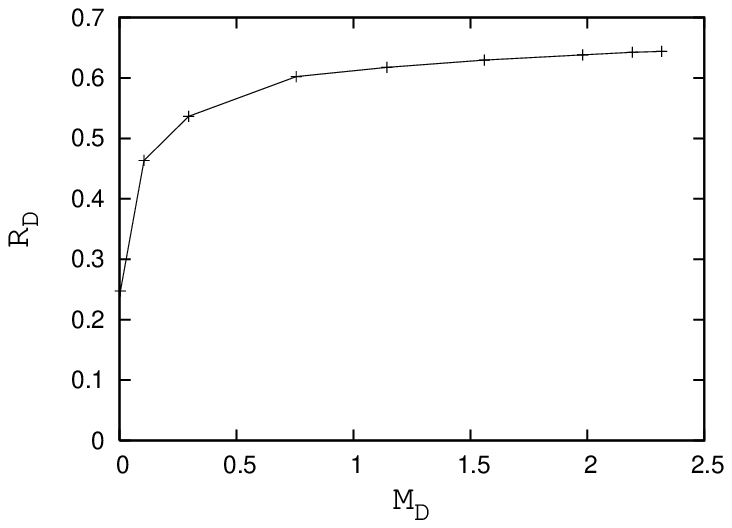}}&
   \subfloat[]{\includegraphics[scale=1.]{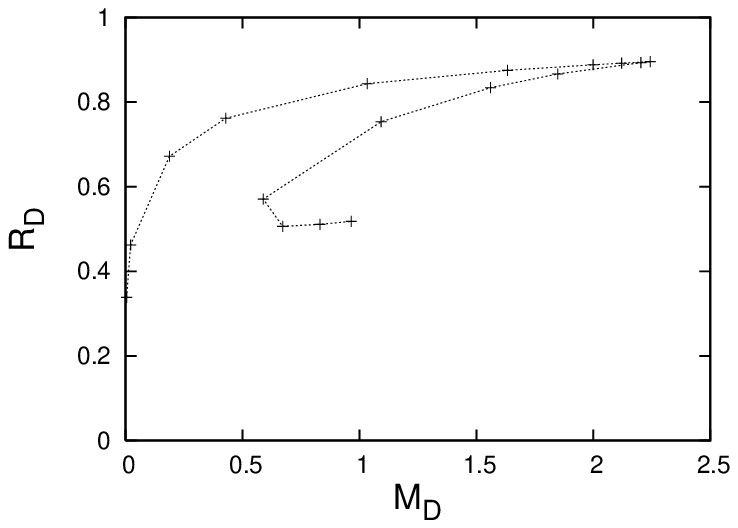}} \\ \\ \\
    \subfloat[]{\includegraphics[scale=1.]{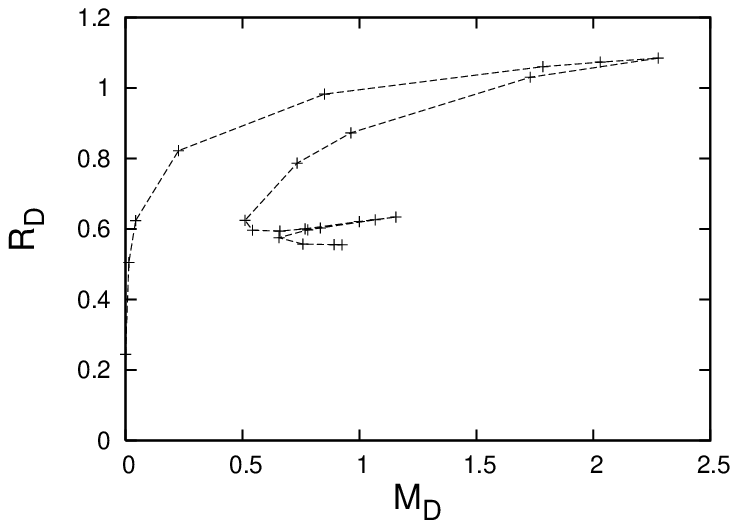}}
           \end{tabular}
    \caption{Mass-radius relations with $E_{Fmax} = 20\,m_{e}c^{2}$ for (a) one-level system, (b) two-level system, (c) three-level system. Here $M_D$ is the mass of the star in units of $M_{\odot}$ and $R_D$ is the radius of the star in units of $10^{8}$ cm (the solid, dotted and dashed lines have the same meaning as in Figure 1).}
    \label{mr1}
  \end{center}
\end{figure}

Figure 2(a) shows the mass-radius relation for the one-level system ($B_{D} = 199.5$). We see that initially as $\rho_{c}$ increases both the mass and radius increase and then at higher central densities the radius becomes nearly independent of the mass (we will be explaining later in this section as to why such a trend is observed). We note that the last point on this curve has a mass $\sim$ 2.3 $M_{\odot}$ and radius $\sim$ $6.4 \times 10^{7}$ cm, which corresponds to the maximum density point of the solid curve in Figure 1(b). This denotes the density ($ \sim 1.16 \times 10^{10}$ gm/cc) at which the ground Landau level is completely filled. Thus a star with this $\rho_{c}$ and a magnetic field strength of $B = 199.5 B_{c} = 8.81 \times 10^{15}$ G has a mass greater than the Chandrasekhar limit ($\sim$ 1.44 $M_{\odot}$ for $\mu_{e} = 2$). 

Figure 2(b) shows the mass-radius relation for the two-level system ($B_{D} = 99.75$). In this case, a maximum mass $\sim 2.3 M_{\odot}$ is reached at a radius $\sim 8.9 \times 10^{7}$ cm for $\rho_{c} \sim 4.0 \times 10^{9}$ gm/cc, in the same way as in Figure 2(a) when the ground Landau level is completely filled. After which there is a turning point in the curve, from where the mass starts decreasing. This turning point corresponds to the kink in the corresponding equation of state (dotted curve in Figure 1(b)). During the transition (when $ 4.0 \times 10^{9} \textrm{ gm/cc} < \rho_{c} \lesssim 8.0 \times 10^{9} $ gm/cc) from ground to first Landau level, the radius and mass both decrease with increasing $\rho_{c}$. Then there is a brief range of densities ($ 8.0 \times 10^{9} \textrm{ gm/cc} < \rho_{c} \lesssim 1.2 \times 10^{10} $ gm/cc) where the radius decreases as the mass remains fairly constant and ultimately at very high densities ($ 1.2 \times 10^{9} \textrm{ gm/cc} < \rho_{c} \lesssim 1.5 \times 10^{10} $ gm/cc) the the radius is again nearly independent of the mass, as in the uppermost branch.

Figure 2(c) shows the mass-radius relation for the three-level system ($B_{D} = 66.5$). Here we see two turning points, denoted by the decrease of both mass and radius, which correspond to the two kinks in the corresponding equation of state (dashed curve in Figure 1(b)). The maximum mass at the first turning point $\sim 2.3 M_{\odot}$ is reached at a radius $\sim 1.1 \times 10^{8}$ cm for $\rho_{c} \sim 2.2 \times 10^{9}$ gm/cc (the first kink in the equation of state). The mass at the second turning point is $\sim 1.2 M_{\odot}$ which has a radius $\sim 6.3 \times 10^{7}$ cm for $\rho_{c} \sim 7.6 \times 10^{9}$ gm/cc (the second kink in the equation of state). Just after either of the turning points denoted by the decrease of radius and mass both, briefly the radius decreases as the mass remains almost constant and finally the radius becomes nearly independent of mass. The maximum mass $\sim 2.3 M_{\odot}$ occurs at the central density where the ground Landau level is completely filled and transition to the first Landau level is about to start. We observe that this density follows $\rho_{c}(\rm{three-level}) < \rho_{c}(\rm{two-level}) < \rho_{c}(\rm{one-level})$ as is also seen from the positions of the kinks in Figure 1. 

In order to explain this behavior in further detail, we resort to the Lane-Emden relations (\ref{rvsrho}), (\ref{mvsrho}) and (\ref{rvsm}). From (\ref{rvsrho}) we note that  if $\Gamma > 2$, then $R$ increases with $\rho_{c}$ and if $\Gamma = 2$, then $R$ is independent of $\rho_{c}$. From (\ref{mvsrho}) we note that if $\Gamma > 4/3$, then $M$ increases with $\rho_{c}$ and if $\Gamma = 4/3$, then $M$ is independent of $\rho_{c}$. Finally from (\ref{rvsm}) we note that if $\Gamma > 2$, then $R$ increases with $M$, if $\Gamma = 2$, then $R$ is independent of $M$ and if $\Gamma = 4/3$, then $M$ is independent of $R$. This is exactly what is observed in Figure 2.

\begin{table}
\vskip0.2cm {\centerline{\large Table 2}}{\begin{center} Parameters for the fitting function of the equation of state shown 
in Figure 1(d) and the corresponding mass-radius relation shown in Figure 2(c).\end{center}}

\begin{center}
\begin{tabular}{|c|c|c|c|c|}

\hline
$\rho_{D}$ in units of $2 \times 10^{9}$gm/cc & $\Gamma$ & $n = \frac{1}{\Gamma - 1}$ & $K$ in CGS units & $R \propto M^{\frac{1-n}{3-n}}$  \\
\hline

0 $-$ 0.096 & 2.9 & 0.526 & 4.055 & $R \propto M^{0.192}$ \\
0.096 $-$ 0.307 & 2.4 & 0.714 & 1.284 & $R \propto M^{0.125}$ \\ 
0.307 $-$ 1.128 & 2.1 & 0.909 & 0.914 & $R \propto M^{0.044}$ \\
1.128 $-$ 2.117 & 0.35 & -1.538 & 1.105 & unstable \\
2.117 $-$ 2.956 & 4/3 & 3 & 0.522 & $M \propto R^{0}$ \\
2.956 $-$ 3.842 & 2.0 & 1.0 & 0.246 & $R \propto M^{0}$ \\
3.842 $-$ 4.651 & 0.35 & -1.538 & 2.225 & unstable \\
4.651 $-$ 6.116 & 4/3 & 3 & 0.496 & $M \propto R^{0}$ \\
6.116 $-$ 7.37 & 2.0 & 1.0 & 0.142 & $R \propto M^{0}$ \\ \hline

\end{tabular}

\end{center}

\end{table}

Let us look again at the mass-radius relation corresponding to Figure 2(c) (also see Table 2). At very low densities $\Gamma$ is $\sim 3$ (which is the case for non-relativistic electrons in the ground Landau level; $P_{e}(= P) \propto \rho^{3}$; see \cite{abrahams}) and the value of $\Gamma$ keeps decreasing with increasing density.  Up to the first turning point density $\rho_{c} \sim 2.2 \times 10^{9}$gm/cc, both the radius and mass keep increasing with $\rho_{c}$ and then the radius becomes nearly independent of mass when $\Gamma \sim 2$. Then $\Gamma$ suddenly drops to the small value $\sim 0.35$ which marks the onset of the transition from ground Landau level to first Landau level. In this region the pressure becomes independent of density, revealing an unstable zone in the equation of state (see detailed discussion in \S4.5). As the density increases further, $\Gamma$ approaches the relativistic value of $4/3$. In this regime we see that the radius decreases slightly as the mass does not change significantly, as is also true for the mass-radius relation in the classical non-magnetic case for $\Gamma=4/3$ (see Figure 4(b)). Next $\Gamma$ takes up a value of $2$ and the radius again becomes nearly independent of mass till it reaches the second turning point at density $\rho_{c} \sim 7.6 \times 10^{9}$ gm/cc. Again during the transition from the first Landau level to the second, $\Gamma$ drops to $0.35$, followed by values of $4/3$ and $2$, which have the same explanations as stated above.

We also observe that for a certain range of masses in the two-level and three-level systems, it is possible to have multiple values of the radius for a given value of mass. For instance, looking at the two-level system in Figure 2(b) in the range $\sim 0.6-1.0$ $M_{\odot}$, we observe that the same value of mass corresponds to three different values of the radius. Let us call them $R_{1}$, $R_{2}$ and $R_{3}$, such that $R_{1} > R_{2} > R_{3}$ and $\rho_{c}(R_{1}) < \rho_{c}(R_{2}) < \rho_{c}(R_{3})$. To explain why we observe such a behavior, we recall that $B_{D}$ of the system is such that $\nu_{m}$ is fixed for a given value of $E_{Fmax}$ (see Table 1 for the values). In Figure 2(b), $B_{D}=99.75$ ensures that the system can have at the most two Landau levels (ground and first), but to what extent they will be filled depends on the Fermi energy of the electrons ($\leq E_{Fmax}$). For low central densities (i.e. low Fermi energy) the electrons occupy only the ground Landau level. In order for the electrons to start occupying the first Landau level, the central density of the star must be adequately high. Thus, based on previous discussion for the mass-radius curves in Figure 2, it is possible that the same mass of a star corresponds to more than one radius, depending on the Landau level occupancy. Now, $R_{1}$ lies on the branch of the mass-radius relation which corresponds to the stars in which the electrons occupy only the ground Landau level, while $R_{3}$ lies on the branch which corresponds to the electrons occupying the first Landau level (the ground level being already filled). $R_{2}$ lies on the unstable branch of the mass-radius relation (see discussion in \S4.5) when the electrons are in the transition mode form the ground to the first Landau level. 
The three-level system in Figure 2(c) can also be explained likewise, except that it has two additional possible radii corresponding to the same mass due to the presence of the second Landau level.  In the one-level system in Figure 2(a), multiple values of radius are not observed because the magnetic field ($B_{D} = 199.5$) is such that the electrons can occupy only the ground Landau level.

\captionsetup[subfigure]{position=top}
\begin{figure}
  \begin{center}
    \begin{tabular}{ll}
 \subfloat[]{\includegraphics[scale=1.0]{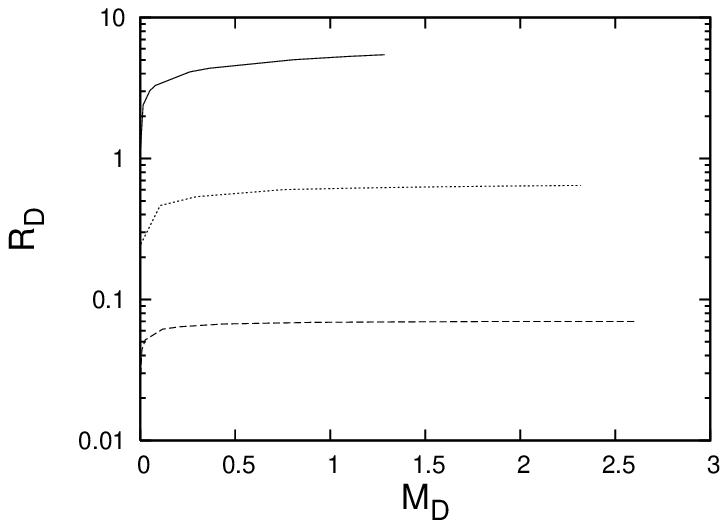}}&
   \subfloat[]{\includegraphics[scale=1.0]{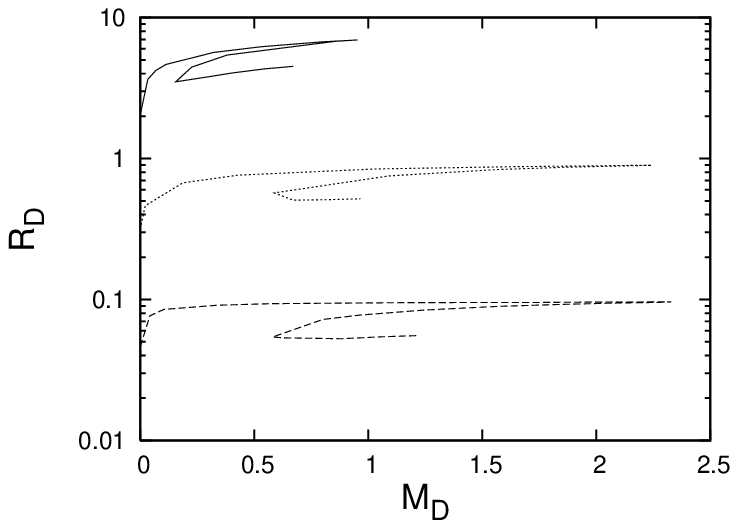}} \\ \\ \\
    \subfloat[]{\includegraphics[scale=1.0]{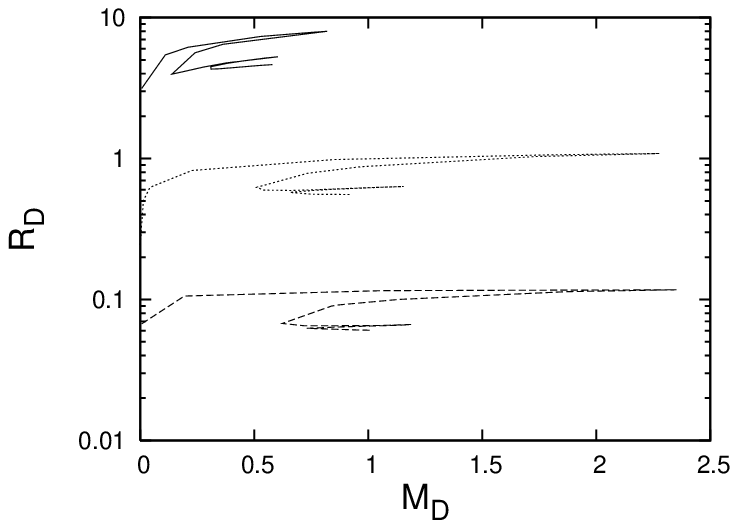}} 
       \end{tabular}
    \caption{Comparison of the mass-radius relations, for (a) one-level system, (b) two-level system, (c) three-level system, when the solid, dotted and dashed lines represent $E_{Fmax} = 2\,m_{e}c^{2}, 20\,m_{e}c^{2}$ and $200\,m_{e}c^{2}$ respectively. In each of the three panels the $y$-axis is in log scale. See Table 1 for details.}
   \label{mr2}
  \end{center}
\end{figure}

Finally we come to Figure 3. For each of the Figures (a), (b) and (c) the value of $E_{Fmax}$ increases from the top to the bottom curve and we see that the overall radius decreases from the top to the bottom curve. We also note that an increase in $E_{Fmax}$ corresponds to an increase in the magnetic field for a fixed $\nu_{m}$, which means that as the magnetic field strength increases the degenerate stars become more and more compact in size.
The same trend has been observed in the case of neutron stars with a high magnetic field \cite{monica}, \cite{ferrer}, \cite{dex} (see discussion in \S4.2.) Interestingly, as seen from Figure 3(a), for $E_{Fmax}=200\,m_{e}c^{2}$ the maximum mass of the star is even higher than that of $E_{Fmax} = 20\,m_{e}c^{2}$, which is $\sim 2.6M_{\odot}$. Also to be noted is the fact that for a given $\nu_{m}$, the curve corresponding to $E_{Fmax} = 20\,m_{e}c^{2}$ covers almost the same range in mass as that of the $E_{Fmax} = 200\,m_{e}c^{2}$ curve, while the curve for $E_{Fmax} = 2\,m_{e}c^{2}$ covers a considerably smaller range in mass. The reason behind this saturation at higher $E_{Fmax}$ may be due to the fact that a Fermi energy $\sim 2\,m_{e}c^{2}$ corresponds to very low density such that the electrons are at the most only mildly relativistic. By the time Fermi energy reaches a value of about $20\,m_{e}c^{2}$ (which corresponds to high density), the electrons have become highly relativistic, giving rise to denser, more massive stars and hence further increase of $E_{Fmax}$ could not bring any new effect in the system.

\section{Discussions}

\subsection{Timescale of decay of magnetic field}

\indent \indent Generally the magnetic fields inside an electron degenerate star undergo Ohmic decay. The timescale 
for this is given by $t_{ohm}=\frac{4 \pi \sigma_{E} L^{2}}{c^{2}}$, where $\sigma_{E}$ is the electrical conductivity 
and $L$ the length-scale over which the magnetic field changes. Theoretical calculations of Ohmic decay in isolated,
cooling white dwarfs, which are generally known in nature, show that the magnetic field changes little over their 
lifetime \cite{gabriel}, \cite{fontaine}, \cite{wendell}. Cumming \cite{cumming} estimated a lowest order decay 
time as
\begin{equation}
t_{ohm} \approx 10^{10} \,\rm yrs \left(\frac{\rho_{c}}{3\times 10^{6}\, \rm gm/cc}\right)^{1/3}\left(\frac{\rm R}{10^{9}\, \rm cm}\right)^{1/2} \left(\frac{10<\rho>}{\rho_{c}}\right),
\label{ohm}
\end{equation}
where $<\rho>$ is the mean density of the star. 
For the three stars represented in Figure 5(d), above timescale turns out to be $\gtrsim 10^{9}$ years. Thus we 
can say that the magnetic fields of these stars do not decay significantly via Ohmic dissipation in their lifetime. 
However, while for $B \lesssim 10^{11}$ G, Ohmic decay is the dominant phenomenon, for 
$B \sim 10^{12}-10^{13}$ G decay is supposed to take place via Hall drift and
for $B \gtrsim 10^{14}$ G, it undergoes ambipolar diffusion \cite{heyl}. Note that 
all the decay timescales stated above are valid only for normal or non-superfluid matter.
Hence, while they might be unrealistic for neutron stars \cite{db99}, 
these timescales are applicable to white dwarfs since they do not have a
superconducting core \cite{ginz,ostgard}.

Heyl \& Kulkarni \cite{heyl}
examined the consequences of the magnetic field decay in magnetars having surface field $10^{14}$ G to 
$10^{16}$ G by using an
appropriate cooling model and by solving the following decay equation:
\begin{equation}
\frac{dB}{dt} = -B\left(\frac{1}{t_{ohmic}} + \frac{1}{t_{ambip}} + \frac{1}{t_{Hall}}\right).
\end{equation}
The strongly magnetized white dwarfs considered in the present work have
central magnetic field strengths $\sim 10^{15}$ G,
which is comparable to the surface field strengths of magnetars. It is then likely that
the magnetic fields in both these cases would undergo similar decay mechanisms.
The decay of such strong fields is dominated by ambipolar diffusion \cite{heyl}. However,
for a typical initial field strength of $10^{15}$ G, the magnetic field remains nearly
constant up to about $10^{5}$ years and in the next $100$ years its value decreases
by at most an order of magnitude \cite{heyl}.  As discussed in \S4.3 below, it is
the central magnetic field which is crucial for the super-Chandrasekhar mass of the
white dwarfs. Thus applying the above results to these magnetized white
dwarfs, we can conclude that the central field will not decay appreciably for a long
period of time.

An alternate scenario could arise if the magnetized white dwarfs are accreting. In
this case, the heat generated due to accretion decreases the electrical conductivity
of the surface of the star, causing a faster decay of the (surface) magnetic field
due to a reduced Ohmic decay timescale. However, as the mass of the star increases,
it becomes more compact and the current carrying accreted material is pushed deeper
into the star. Since conductivity is a steeply increasing function of density, the
higher conductivity of the denser inner region of the star will again slow down
further decay of the magnetic field \cite{konar}.

Since we are working with magnetic field strengths $B>B_{c}$, one might be concerned about the process of 
electron-positron pair creation (via Schwinger process) at the expense of magnetic energy, which might lead 
to a reduction of the field strength. However, Canuto and Chiu \cite{can1}, \cite{can2}, \cite{can3}  and also Daugherty et al. \cite{daugh} showed that it is impossible to have spontaneous pair creation in a magnetic field alone, irrespective of its strength. Now, from Maxwell's equations in a steady state, we have
\begin{equation}
\nabla \times {\bf B} = \frac{4\pi}{c}{\bf j}
\label{amp}
\end{equation}
and Ohm's law states
\begin{equation}
{\bf E}=\frac{\bf j}{\sigma_{E}},
\label{ohm}
\end{equation}
when ${\bf E}$ is the electric field and ${\bf j}$ the current density. Thus for an electric field to be generated, the magnetic field must vary with space as seen from equation (\ref{amp}). However, in this work we have chosen a constant magnetic field, which leads to ${\bf j}=0$ and hence ${\bf E}=0$ from equation (\ref{ohm}). Now the main effect of the magnetic field, which is to give rise to a mass exceeding the Chandrasekhar limit, is restricted to the high density region, where the field remains essentially constant. Thus our choice of a constant magnetic field is justified (see \S4.3 for detailed discussion). Moreover since the magnetic field in the degenerate star is generated due to the flux freezing phenomenon, it incorporates the fact that the conductivity $\sigma_{E}$ is very large. Thus even if the magnetic field is highly inhomogeneous, from equation (\ref{ohm}) we see that the electric field generated would again be negligible. Thus the electron degenerate stars in this work are magnetically dominated systems, i.e., both the magnetic field strength is very high ($B>B_{c}$) and the electric field strength is negligible. Recently, Jones \cite{jones} calculated the cross-section for photon-induced pair creation in very high magnetic fields and has arrived at the result that, there is a rapid decrease of the pair creation cross-section at $B>B_{c}$. Hence one can ignore the effect of pair creation in reducing the magnetic field strength in these stars.

\subsection{Comparison with Chandrasekhar's results}

\captionsetup[subfigure]{position=top}
\begin{figure}
  \begin{center}
    \begin{tabular}{ll}
 \subfloat[]{\includegraphics[scale=0.75]{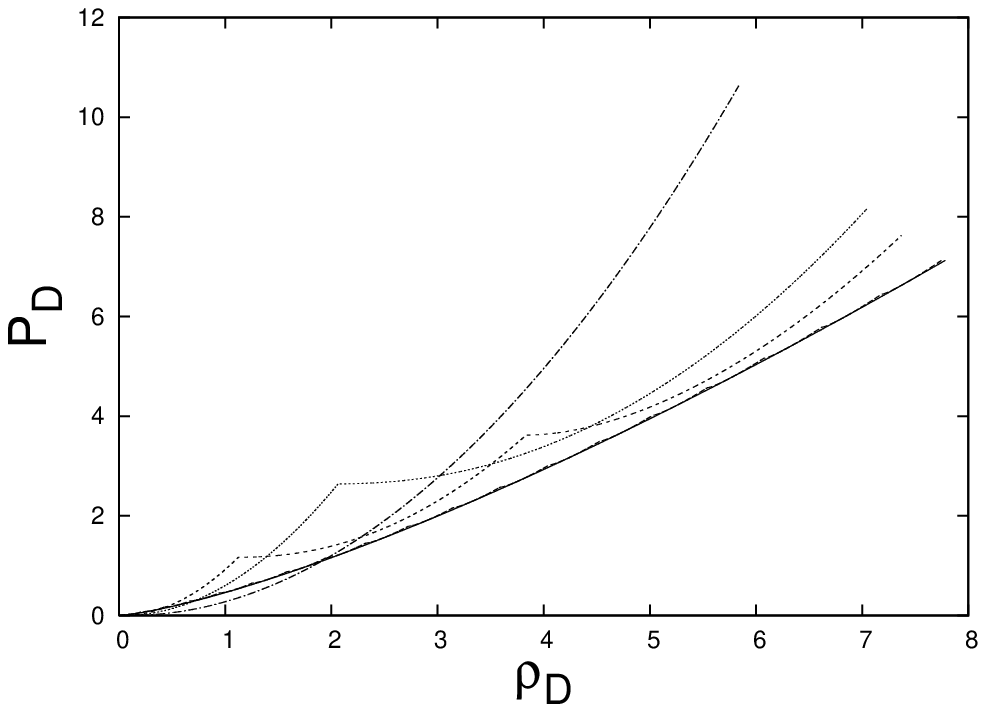}}&
   \subfloat[]{\includegraphics[scale=0.75]{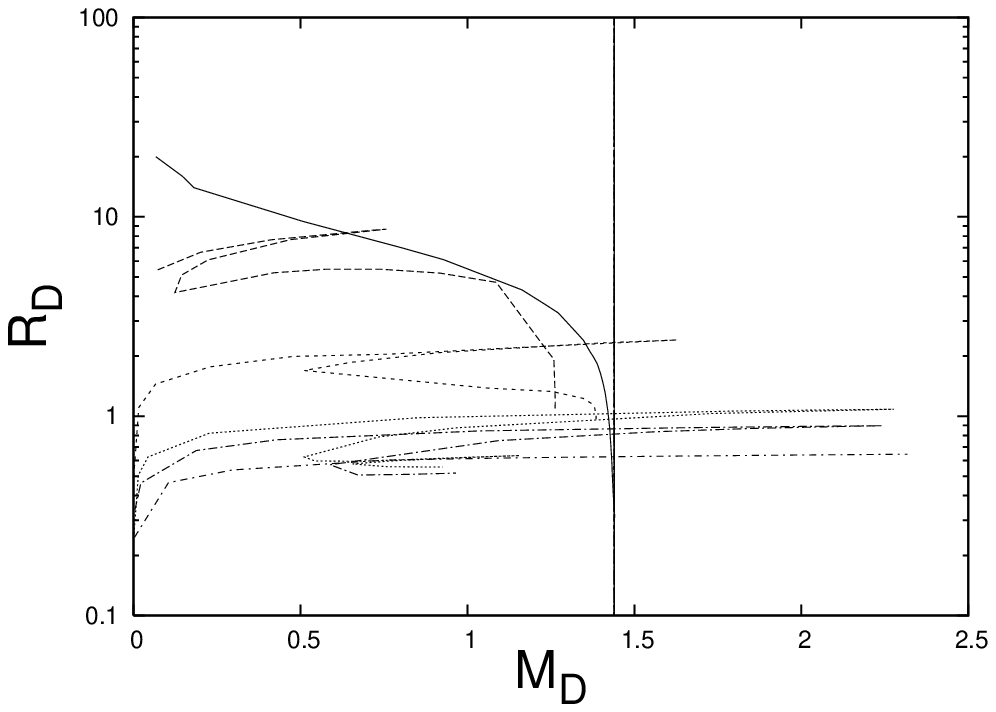}} \\ \\ \\
 \subfloat[]{\includegraphics[scale=0.75]{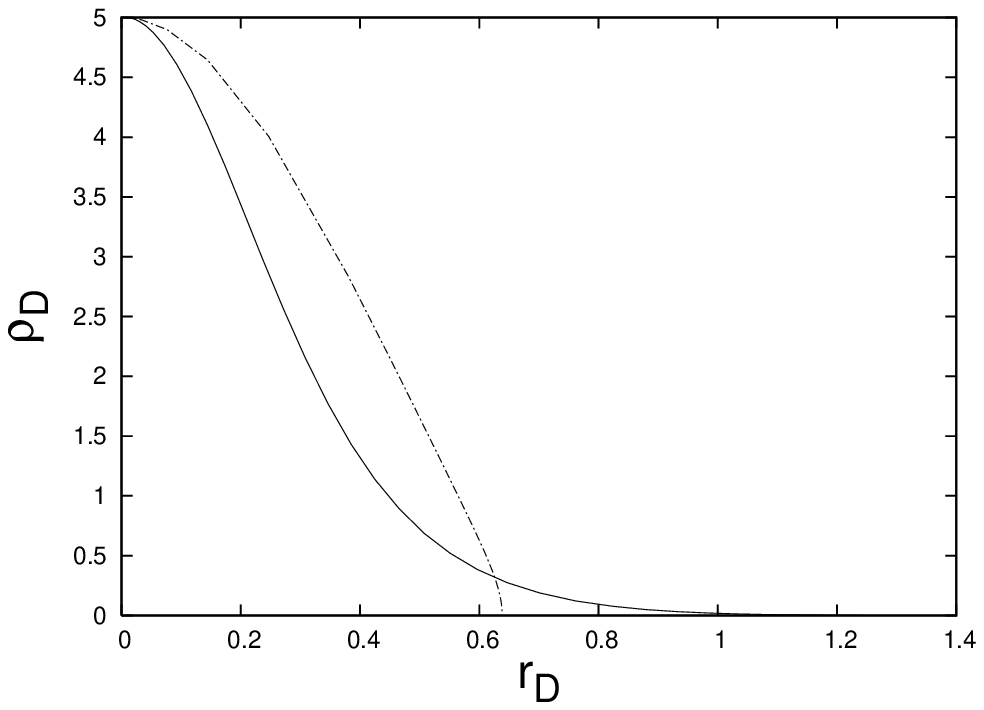}} 
           \end{tabular}
    \caption{Comparison with Chandrasekhar's non-magnetic results for $E_{Fmax} = 20\,m_{e}c^{2}$. (a) Equations of state - the solid line represents Chandrasekhar's equation of state. The dot-dashed, dotted and dashed lines represent the one-level ($\nu_{m}=1$), two-level ($\nu_{m}=2$) and three-level ($\nu_{m}=3$) systems respectively. The equation of state for $\nu_{m} = 20$ is also shown, which appears as a series of kinks on top of the solid line. (b) Mass-radius relations - the vertical line marks the 1.44$M_{\odot}$ limit and the solid line represents Chandrasekhar's mass-radius relation. From top to bottom the other lines represent the cases for $\nu_{m} = 500\,, 20\,, 3\,, 2$ and $1$ respectively (the $y$-axis is in log scale). (c) Density as a function of radius inside a non-magnetized star (solid line) and a star having $\nu_{m}=1$ (dot-dashed line), both having the same central density $\rho_{c}=5$ in units of $2 \times 10^{9}$ gm/cc.} 
   \label{nonmagnetic}
  \end{center}
\end{figure}

\indent \indent In Figure 4 we put together our results along with Chandrasekhar's result for non-magnetic white dwarfs to obtain a more complete picture. Figure 4(a) shows the equation of state obtained by Chandrasekhar and those corresponding to the one ($\nu_{m}=1$), two ($\nu_{m}=2$) and three ($\nu_{m}=3$) level systems (same as in Figure 1(b)). Interestingly, the equation of state for $\nu_{m}=20$ almost grazes Chandrasekhar's equation of state, except the appearance of a series of kinks. This clearly shows that as the magnetic field strength decreases, or equivalently as the maximum number of occupied Landau levels increases, the equation of state approaches Chandrasekhar's equation of state.

Figure 4(b) represents the mass-radius relations corresponding to the equations of state shown in Figure 4(a). The mass-radius relation for $\nu_{m}=500$ (500 Landau levels) is also shown. We observe that as the magnetic field strength decreases or as the maximum number of occupied Landau levels increases, the mass-radius relation approaches the non-magnetic relation and one recovers Chandrasekhar's mass limit. 

Figure 4(b) also shows that the stars of a given mass become more and more compact in size as the magnetic field strength increases. In order to explain this behavior we resort to Figures 4(a) and (c). Let us for simplicity look at Chandrasekhar's equation of state (solid line) and that of the one-level system (dot-dashed line), which corresponds to $B=199.5B_{c}$, in Figure 4(a). We notice that at low densities, the dot-dashed line lies below the solid line and at higher densities, $\rho_{D}\gtrsim 2$, the dot-dashed line lies above the solid line. In other words, the equation of state for the one-level system is softer than Chandrasekhar's equation of state at low densities, which means that the pressure does not rise with density as rapidly as that in Chandrasekhar's case. This trend reverses at higher densities and the equation of state for the one-level case becomes stiffer than that of Chandrasekhar's. Now, matter with a softer equation of state is less efficient in counteracting gravity and hence stars made out of such matter will be more compact in size. Keeping this in mind we now look at Figure 4(c). Figure 4(c) shows the variation of density with radius, within a non-magnetized star and a magnetized star with $\nu_{m}=1$ (one-level), both having the same central density ($\rho_{c}=5$). We mention here that for ease of explanation we have chosen a high central density such that the equations of state for both the stars cover almost the entire range of density. We observe that for a given radius, density for the magnetized star is higher than that of the non-magnetized star for a large range, from $\rho_{D}=\rho_{c}=5$ to $\rho_{D} \sim 0.3$. But at very low densities, $\rho_{D}<0.3$, the density of the magnetized star sharply falls to zero due to smaller pressure, leading to a smaller star ($R=6.4\times 10^{7}$ cm) than in the non-magnetized case ($R=1.3\times 10^{8}$ cm). From the analysis of the equations of state for these two stars, we can say that if the equation of state of the magnetized star would have remained stiffer than the non-magnetized star throughout, then the magnetized star would have a larger radius, as in this case the pressure would be more efficient in counteracting the gravitational collapse of the star. But this is not true. At very low densities the equation of state suddenly becomes softer for the magnetized star and the pressure is not able to counteract gravity efficiently, causing the star to collapse rapidly. Hence the density goes to zero very rapidly, causing the star to have a (much) smaller radius. 

We also note from Figure 4(b) that as the magnetic field strength increases, the probability that the stars will have masses exceeding the Chandrasekhar limit increases. From Figure 4(c) we see that the density of the magnetized star is much higher than that of the non-magnetized star, except for a very small range of low densities. Thus calculating the total mass of the stars from equation (\ref{mass}), we obtain a much higher value for the magnetized star.

\subsection{Choice of constant magnetic field}

\captionsetup[subfigure]{position=top}
\begin{figure}
  \begin{center}
    \begin{tabular}{ll}
 \subfloat[]{\includegraphics[scale=0.8]{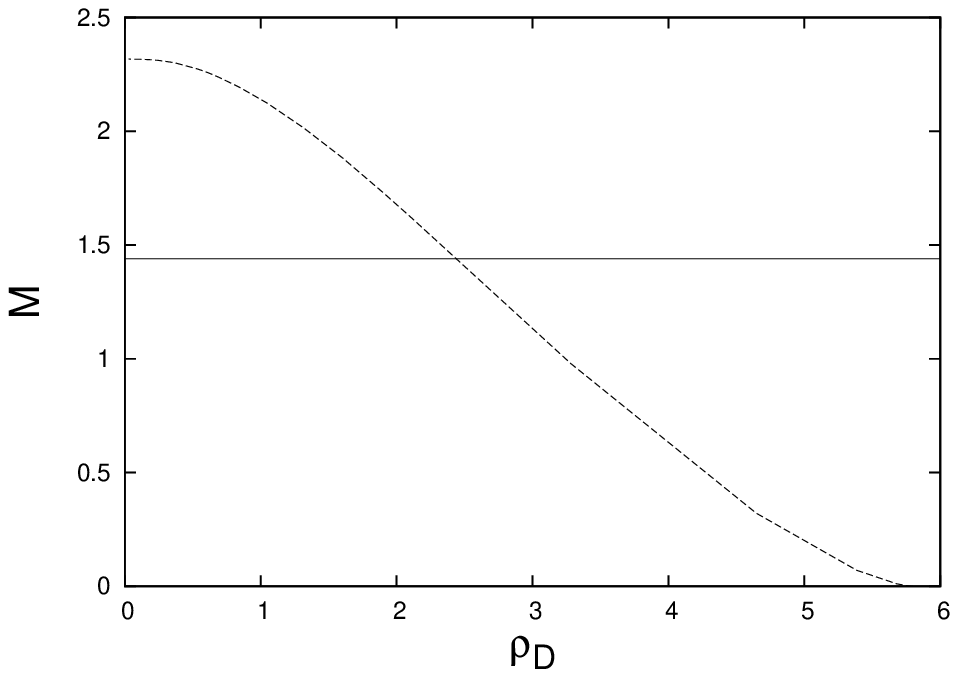}}&
   \subfloat[]{\includegraphics[scale=0.8]{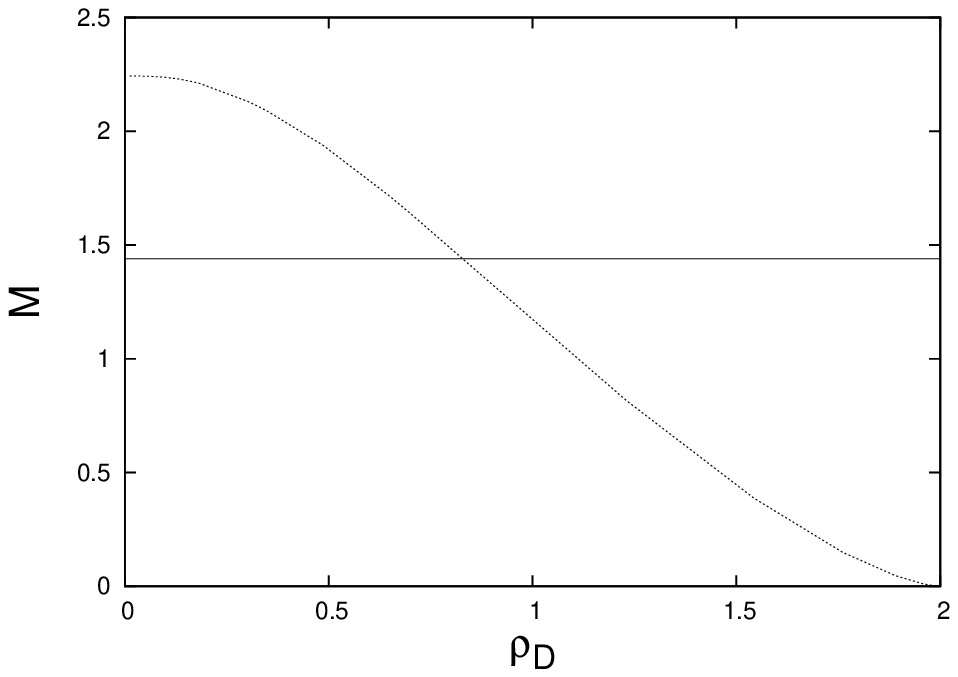}} \\ \\ \\
    \subfloat[]{\includegraphics[scale=0.8]{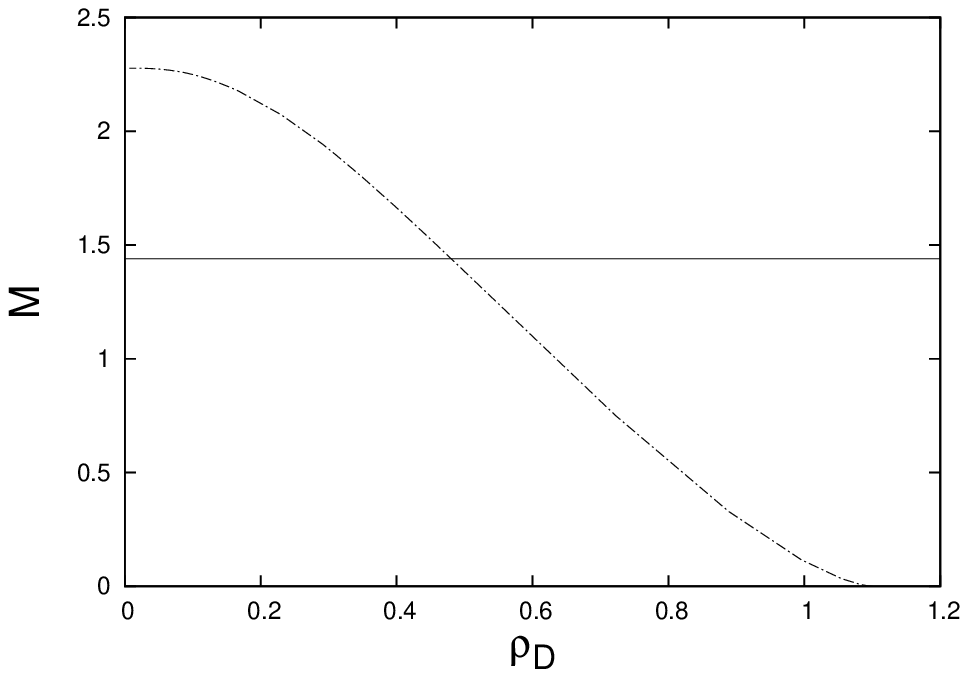}}& 
 \subfloat[]{\includegraphics[scale=0.8]{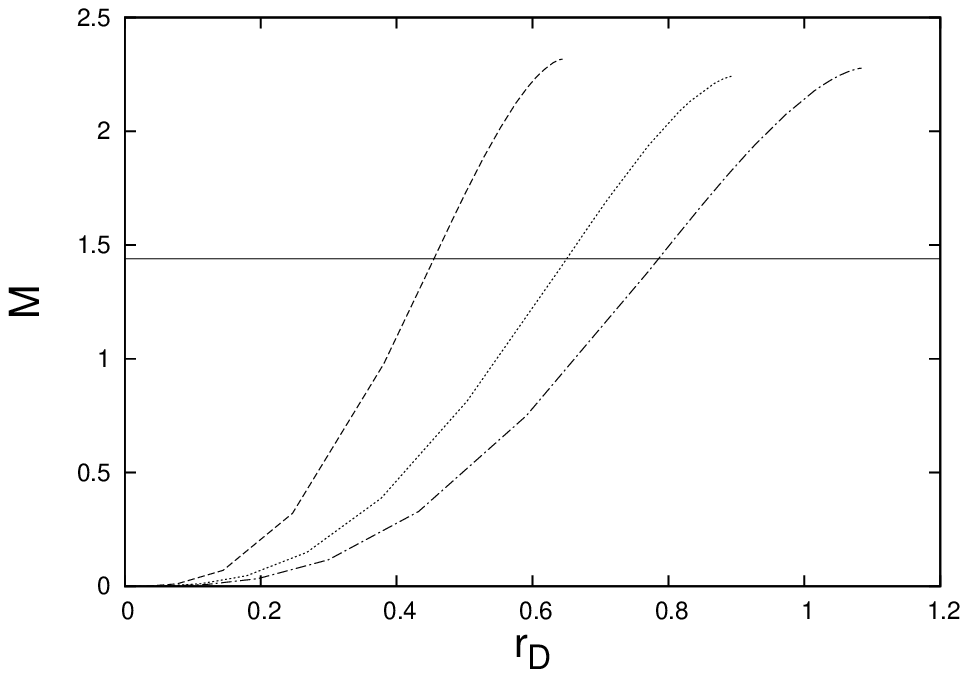}} \\
       \end{tabular}
    \caption{Mass as a function of density  within an electron degenerate star with $B_{D}$ (a) 199.5 (dashed line), (b) 99.75 (dotted line), (c) 66.5 (dot-dashed line), for $E_{Fmax} = 20\,m_{e}c^{2}$. (d) Variation of mass as a function of radius within the stars in (a), (b) and (c). The horizontal line indicates the 1.44$M_{\odot}$ limit. The radius $r_{D}$ is in units of $10^{8}$ cm. See Table 1 for details.}
   \label{mrho}
  \end{center}
\end{figure}

\indent \indent Figures 5(a), (b) and (c) show the variation of mass as a function of density within a magnetized electron degenerate star for three different magnetic field strengths. In all the cases we note that, by the time the density falls to about half the value of the central density, the mass has increased significantly, crossing the Chandrasekhar limit (indicated by the horizontal line) soon after. Hence, although we have considered a constant magnetic field, their effect is restricted to the high density regime, where the field remains essentially constant in reality. Hence we can also interpret this constant magnetic field as the central magnetic field of the star. This would be more clear from the description of the variation of magnetic field given by \cite{monica}, \cite{bandyo}, which show that an inhomogeneous magnetic profile in a compact star could be such that the magnetic field is nearly constant throughout most of the star and then gradually falls off close to the surface (see Figure 5(b) in \cite{monica}). Thus choosing an inhomogeneous magnetic profile would not affect our main finding that the Chandrasekhar mass limit can be exceeded for high magnetic field strengths.

\subsection{Choice of non-relativistic equation of hydrostatic equilibrium}

\indent \indent Figure 5(d) shows the variation of mass as a function of radius within the stars represented in Figures 5(a), (b) and (c). All these stars have a total mass $\sim 2.3M_{\odot}$ and hence their Schwarzschild radius $R_{g}=\frac{2GM}{c^{2}}=6.8\times10^{5}$ cm. Now, general relativistic effects usually start becoming important at a radius $\lesssim 10 R_{g}$, i.e. $r_{D} \lesssim 0.068$ for the above mentioned stars. However, from Figure 5(d) we see that for all the three stars the contribution to the mass from a radius $< 10R_{g}$, i.e. the central region, is negligible. Contribution to the mass rather effectively starts from a radius $r_{D} \gtrsim 0.13$ ($\sim 20 R_{g}$) and the Chandrasekhar limit is crossed at $R_{g} \sim 66,\, 95$ and $118$, for the stars represented by the dashed, dotted and dot-dashed lines, respectively. Thus significant contribution to the total mass of the stars comes from a region well beyond the regime of general relativity. Hence one need not necessarily consider the Tolman-Oppenheimer-Volkoff equation and our choice of the non-relativistic equation of equilibrium, equation (\ref{diff}), is justified.

\subsection{Unstable branch of the mass-radius relations}

\indent \indent Figure 6(a) shows the variation of density as a function of radius inside the degenerate stars having $E_{Fmax} = 20\,m_{e}c^{2}$ and $B=99.75B_{c}$. If we look at the equation of state for the two-level system with $E_{Fmax} = 20\,m_{e}c^{2}$, i.e. the dotted line in Figure 1(b), we see that both $\rho_{D}=1.5$ and $2$ lie below the density at which transition takes place from ground Landau level to first Landau level. For stars with these central densities, the pressure rises monotonically with density throughout and one forms a stable star. From the corresponding curves (solid and dashed) in Figure 6(b) also we note that the sound speed ($c_{s}$) varies smoothly with radius, reaching a maximum value at the center. Both the mass and radius of these stars increase with central density and hence they lie on the uppermost branch of the mass-radius relation in Figure 2(b).

\captionsetup[subfigure]{position=top}
\begin{figure}[h]
  \begin{center}
\begin{tabular}{ll}
 \subfloat[]{\includegraphics[scale=0.75]{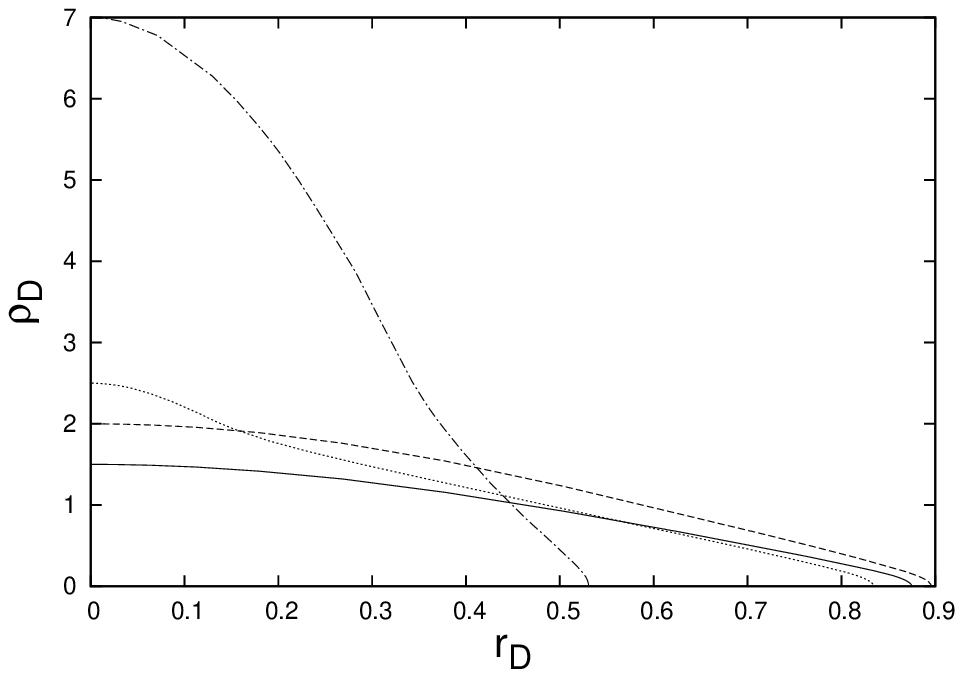}}&
   \subfloat[]{\includegraphics[scale=0.75]{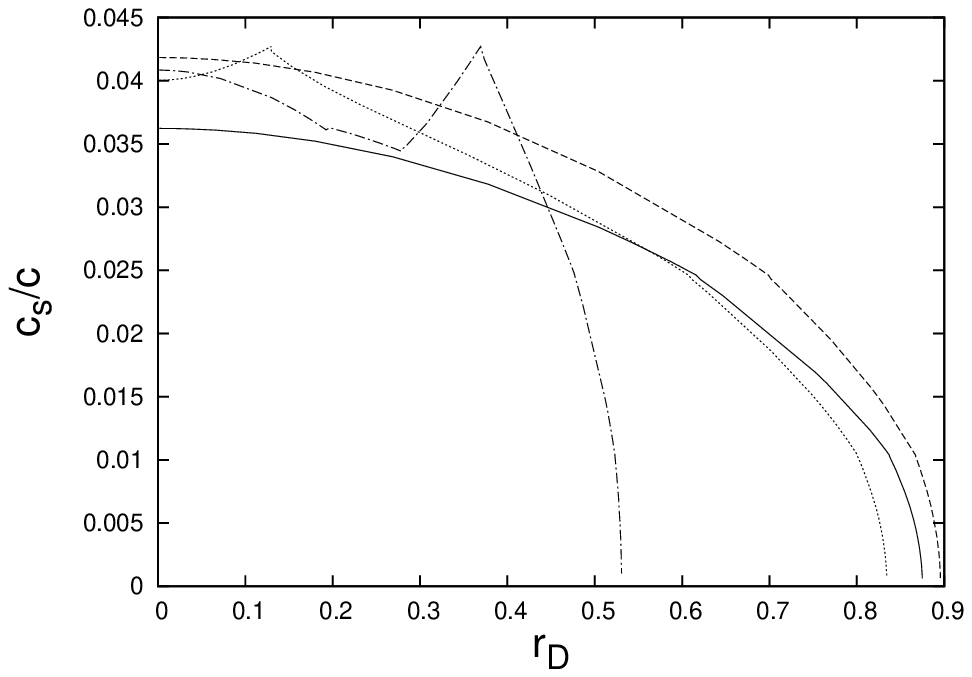}} \\ 
\end{tabular}
    \caption{(a) Density and (b) sound speed in units of $c$, as a function of radius, inside the stars having $E_{Fmax} = 20\,m_{e}c^{2}$ and $B=99.75B_{c}$. The solid, dashed, dotted and dot-dashed lines represent the stars with central densities 1.5, 2, 2.5 and 7, in units of $2 \times 10^{9}$ gm/cc, respectively.}
    \label{prho}
  \end{center}
\end{figure}

Now $\rho_{D}=2.5$ lies on the plateau following the kink in the equation of state. A star having this central density will have within it a zone where pressure does not steadily increase with density, but is nearly constant. Thus such a star will tend to collapse faster under gravity and hence will have a smaller radius. From the dotted line in Figure 6(a) we see that the density of such a star falls steeply to zero compared to the $\rho_{D}=2$ case indicated by the dashed line. The corresponding curve in Figure 6(b) shows a peak in $c_{s}$ at the radius where pressure starts becoming independent of density. Since mass of the star is calculated using equation (\ref{mass}), they have smaller masses too due to smaller size. These stars lie on the middle branch of the corresponding mass-radius relation. Since these stars consist of an unstable zone mentioned above, we argue that they constitute the somewhat unstable branch in the mass-radius relation. 

Finally, let us choose a larger density, say $\rho_{D}=7$, which lies on that portion of the equation of state above the kink, where pressure again rises monotonically with density, but not as steeply as that in the regions around $\rho_{D}=1.5$ and $2$. Although these stars also have an unstable zone inside them, but softer pressure in a large range of density, which together cause them to have a smaller radius, their central density is very high. As a result their mass starts increasing again, as can be inferred from the dot-dashed line in Figure 6(a). The $c_{s}$ again shows a sharp peak indicating the appearance of unstable zone within the star. These stars constitute the bottommost branch of the mass-radius relation.

\subsection{Neglecting Coulomb interactions}

\indent \indent The distance between nuclei in highly magnetized electron degenerate stars could be as small as 50 Fermi. The Coulomb repulsion energy between electrons $\frac{e^{2}}{r}$, at the separation of 50 Fermi, is of the order of $4\times10^{-8}$ ergs which is quite less than the rest mass energy of an electron $m_{e}c^{2} \sim 8\times10^{-7}$ ergs. There can be Coulomb interaction between electrons and the ions, which is given by $\frac{Ze^{2}}{r}$. Commonly electron degenerate stars consist of helium, carbon, oxygen, etcetera, so that $Z$ can have a value of $10$ at the most. Hence, the Coulomb interaction energy would still be less than, or at most the same order as, the rest mass energy. Thus we can neglect the effects of Coulomb interaction for the present purpose.

\subsection{Anisotropy in pressure due to strong magnetic field}

\indent \indent The strong magnetic field causes the pressure to become anisotropic \cite{ferrer}, \cite{khalilov}, \cite{paulucci}. The total energy momentum tensor due to both matter and magnetic field is to be given by
\begin{equation}
T^{\mu \nu} = T^{\mu \nu}_{m} + T^{\mu \nu}_{f},
\end{equation}
where,
\begin{equation}
T^{\mu \nu}_{m} = \epsilon_{m}u^{\mu}u^{\nu} - P_{m}(g^{\mu \nu} - u^{\mu}u^{\nu})
\end{equation}
and
\begin{equation}
T^{\mu \nu}_{f} = \frac{B^{2}}{4\pi}(u^{\mu}u^{\nu} - \frac{1}{2}g^{\mu \nu}) - \frac{B^{\mu}B^{\nu}}{4\pi},
\label{tmunu}
\end{equation}
when, $\epsilon_{m} = \epsilon_{e}$ is the matter energy density given by equation (\ref{endensity}) and $P_{m} = P_{e}$ is the matter pressure given by equation (\ref{pressure}). The first term in equation (\ref{tmunu}) is equivalent to magnetic pressure, while the second term gives rise to the magnetic tension. If $B$ is along the $z$-axis, then we have
\begin{equation}
T^{\mu \nu}_{f} = 
\begin{bmatrix}
\frac{B^{2}}{8\pi} & 0 & 0 & 0 \\
0 & \frac{B^{2}}{8\pi} & 0 & 0 \\
0 & 0 & \frac{B^{2}}{8\pi} & 0 \\
0 & 0 & 0 & -\frac{B^{2}}{8\pi} 
\end{bmatrix}.
\end{equation}
Thus we see that pressure becomes anisotropic. The total pressure in the perpendicular direction to the magnetic field is given by
\begin{equation}
P_{\bot} = P_{m} + \frac{B^{2}}{8\pi}
\end{equation}
and that in the parallel direction to the magnetic field is given by
\begin{equation}
P_{\|} = P_{m} - \frac{B^{2}}{8\pi}.
\end{equation}
Now the parallel pressure becomes negative if the magnetic pressure exceeds the fluid pressure. In order to understand this effect, we write the component of $T^{\mu \nu}_{f}$ along the $z$-axis as 
\begin{equation}
T^{zz}_{f} = \frac{B^{2}}{8\pi} - \frac{B^{2}}{4\pi}.
\end{equation}
The second term $- B^{2}/4\pi$ corresponds to an excess negative pressure or tension along the direction to the magnetic field. Thus the total energy momentum tensor can be written as
\begin{equation}
T^{\mu \nu} = 
\begin{bmatrix}
\epsilon_{m} + \frac{B^{2}}{8\pi} & 0 & 0 & 0 \\
0 & P_{m} + \frac{B^{2}}{8\pi} & 0 & 0 \\
0 & 0 & P_{m} + \frac{B^{2}}{8\pi} & 0 \\
0 & 0 & 0 & (P_{m} +\frac{B^{2}}{8\pi}) - \frac{B^{2}}{4\pi}
\end{bmatrix}.
\end{equation}

The strong magnetic field also reveals anisotropy due to magnetization pressure \cite{ferrer}. 
Hence, actually the pressure 
in the perpendicular direction to the magnetic field is given by
\begin{equation}
P_{\bot} = P_{m} + \frac{B^{2}}{8\pi} - \cal M \textit B,
\end{equation}
where $\cal M$ is the magnetization of the system, which is given by
\begin{equation}
{\cal M} = - \frac{\partial \epsilon_{m}}{\partial \textit B}.
\end{equation}
However, for the magnetic fields considered in the present work exhibiting super-Chandrasekhar masses, $B^2/8\pi>>\cal M \textit B$. 
Therefore, we do not include magnetization term in the pressure which does not affect the result practically for the present purpose.

Here we refer to the work by Bocquet et al. \cite{boquet}, which models rotating neutron stars with magnetic fields, by using an extension of the electromagnetic code used by Bonazzola et al. \cite{bonazola}. They observed that the component of the total energy momentum tensor along the symmetry axis becomes negative (equivalent to $T^{zz} < 0$ in our case), since the fluid pressure decreases more rapidly than the magnetic pressure away from the center of the star. This happens because the combined fluid-magnetic medium develops a tension. As a result of this magnetic tension, the star displays a pinch across the symmetry axis and assumes a flattened shape. A similar effect is expected to occur in our work, where the magnetic tension will be responsible for deforming the magnetized white dwarf along the direction to the magnetic field and turns it into a kind of oblate spheroid. Hence, one should be cautious before considering 
equation (\ref{diff}) which is applicable for a spherical star \cite{paulucci}.

However, in this work we consider a constant magnetic field, since our interest is to see the effect of the magnetic field of the central region of the white dwarf, where the field is supposed to be (almost) constant (see \S4.3). Thus even if we use either the parallel or the perpendicular pressure in the hydrostatic equilibrium  equation (\ref{diff}), $B$ does not appear explicitly in the equation ($dB/dr=0$) --- only the gravitational field will be modified due to deformation. Hence, it is still possible to have super-Chandrasekhar mass white dwarfs - only they will be deformed in shape due to the strong field, which might even render a more massive white dwarf (as is discussed in Appendix A).


\section{Summary and Conclusions}

\indent \indent We have studied the effect of high magnetic field on the equation of state of purely electron degenerate matter at zero temperature. In the equation of state, we have considered only the electron degeneracy pressure modified by the strong magnetic field. We have focused on those Landau quantized systems in which the maximum number of Landau level(s) occupied is/are one, two or three, which we have named to be one-level, two-level and three-level system respectively. 

We have found that whenever a lower Landau level is completely filled and the next higher level is to be filled, a kink appears in the equation of state, followed by a plateau - which is a small region where the pressure becomes nearly independent of the density. The one-level system, which has only the ground Landau level filled, has no kink, the two-level system has one kink at the ground to first level transition and the three-level system has two kinks, one at the ground to first and the other at the first to second-level transition. We have studied each of these systems at three maximum Fermi energies $E_{Fmax} = 2\,m_{e}c^{2}$, $20\,m_{e}c^{2}$ and $200\,m_{e}c^{2}$ and obtained the mass-radius relations of the corresponding stars. 

The mass-radius relations show turning point(s), denoted by the decrease of both mass and radius, which correspond(s) to the kink(s) in the equation of state. The most interesting result obtained is that there are possible stars found on the mass-radius relations whose mass exceeds the Chandrasekhar limit. They could be potential magnetized white dwarfs. The maximum mass obtained is about $2.3-2.6 M_{\odot}$ and is seen to occur for various combinations of central density, magnetic field strength and maximum number of occupied Landau levels. Interestingly such super-massive white dwarfs have been suggested to be the most likely progenitors of recently observed Type Ia supernovae \cite{nature}, \cite{scalzo}. Out of the equations of state considered in the present work, the system with the lowest magnetic field which gives rise to this mass is the three-level system with $E_{Fmax} = 20\,m_{e}c^{2}$, $B_{D} = 66.5$ (or $B = 2.94 \times 10^{15}$ G) and the corresponding central density being $2.2 \times 10^{9}$ gm/cc.

The nature of the mass-radius relations is governed by the fact whether the system is one-level, two-level or three-level and is independent of the value of $E_{Fmax}$. However, $E_{Fmax}$ determines how relativistic the system is. For instance, the Chandrasekhar mass limit is not exceeded for a low $E_{Fmax}$ (say = $2\,m_{e}c^{2}$), no matter what the central density is. We have however observed that as $E_{Fmax}$ increases, which corresponds to an increase in the magnetic field strength, the degenerate stars become more compact in size.

As discussed, the minimum magnetic field required to have a $2.3M_{\odot}$ degenerate star is $B = 2.94 \times 10^{15}$ G. The magnetic field of the original star of radius $R_{\odot}$, which collapses into the above degenerate star of radius $\sim 10^{8}$ cm, turns out to be $\sim 6 \times 10^{9}$ G, based on the flux freezing theorem. Existence of such stars is not ruled out \cite{shapiro}. However, the anisotropy in pressure due to strong magnetic field causes a deformation in the white dwarfs which adopt a flattened shape. This effect of flattening leads to more massive white dwarfs, even at relatively lower magnetic field strengths.

One might wonder as to why highly magnetized white dwarfs have not yet been observed. A plausible reason could be that the surface magnetic field is being screened due to some physical processes. For instance, if the white dwarf is in a binary system and is accreting matter from its companion, as is proposed for Type Ia supernovae progenitors, then the plasma that is being deposited on the surface of the star could induce an opposite magnetic moment. This would result in a reduction of the surface field strength. However, the central magnetic field strength, which is presumably unaffected by the above processes, could be several orders of magnitude higher than the surface field. Indeed as seen in \S4.3, it is the central field which is crucial for exceeding the Chandrasekhar mass limit.



\section*{Acknowledgments}

\indent \indent This work was partly supported by the ISRO grant ISRO/RES/2/367/10-11. We would like to thank the anonymous referees, Dipankar Bhattacharya and Efrain J. Ferrer for their useful comments, which have helped us 
greatly in improving this work.

\appendix
\section{Appendix}

\indent \indent In order to estimate the effect of deviation from spherical symmetry due to the magnetic field we have performed a few calculations. If the magnetic field is very strong, then the magnetic tension will flatten the star along the direction to the field (as discussed in \S4.7). If we consider the white dwarf to be an oblate spheroid with the $z$-axis being the symmetry axis, then its equation is given by
\begin{equation}
\frac{r_{eq}^{2}}{a^{2}} + \frac{z^{2}}{c^{2}} = 1,
\end{equation}
where, $r_{eq}^{2} = x^{2} + y^{2}$, the semi-axis $a$ is the equatorial radius of the spheroid and $c$ is the distance from center to the pole along the symmetry axis ($c<a$ for an oblate spheroid).

\noindent The equatorial force balance equation can be written as
\begin{equation}
\frac{1}{\rho} \frac{dP}{dr_{eq}} = -\frac{GM}{r^{2}}\left(\frac{r_{eq}}{r}\right),
\label{Peq}
\end{equation}
where, $r^{2} = r_{eq}^{2} + z^{2}$. The above equation for hydrostatic equilibrium has to be supplemented by an equation to determine the mass of the star. Now,
\begin{equation}
dM = \rho dV
\end{equation}
and the volume element for an oblate spheroid is given by
\begin{equation}
dV = \pi r_{eq}(z)^{2} dz.
\end{equation}
For a simpler visualization, one can also assume a cylindrical geometry and the equation for the mass can be given by
\begin{equation} 
\frac{dM}{dr_{eq}} = 2\pi r_{eq} h \rho,
\label{Meq}
\end{equation}
where $h$ denotes an average height of the cylinder (assuming that the density does not change appreciably with $z$) and is a parameter that quantifies the degree of flattening. At high field strengths, the white dwarf will be more flattened and $h$ will be small. At low field strengths it is likely that the star will be less flattened and $h$ will have a larger value. Keeping this in mind we solved the above equations and obtained the mass-radius relations for some of the cases as shown in Figure \ref{eos}.

The one-level and two-level systems have strong magnetic fields and the corresponding mass-radius relations for the spherical case are denoted by the solid lines in Figures 7(a) and (b). From the dashed lines in Figures 7(a) and (b) we see that the flattened white dwarfs could have much higher masses than the perfectly spherical ones. Interestingly we note from the dashed lines of Figures 7(c) and (d) that even for much lower magnetic field strengths ($B = 4.4\times10^{14}$ G for $\nu_{m}=20$ and $1.8\times10^{13}$ G for $\nu_{m}=500$), the Chandrasekhar mass limit is exceeded even if one includes the corresponding reduced flattening effect - these white dwarfs were sub-Chandrasekhar for the spherical cases. One can also see that the radius of the stars represented by the dashed lines is larger than that represented by the solid lines. One must note here that for the dashed lines this radius is the equatorial radius, which will automatically be larger than the spherical radius as a consequence of the flattening effect.

Hence, the effect of flattening leads to a more massive star. The effect is very similar to the flattening due to centrifugal force in rapidly rotating stars, which are known to be more massive than their slow-rotating counterparts (e.g. \cite{boquet}). These are also shown to have larger mass in presence of high magnetic field. Thus the strong magnetic field is responsible for the increasing the mass of the white dwarfs, while the deformation (or flattening) of the white dwarf due to the field further adds on to the mass. Therefore, our estimate of the mass of the white dwarf, in fact just sets a lower bound. More interestingly, flattening effects due to magnetic field will render super-Chandrasekhar white dwarfs even at a smaller magnetic field - such relatively low magnetized white dwarfs are more probable in nature.

\captionsetup[subfigure]{position=top}
\begin{figure}
  \begin{center}
    \begin{tabular}{ll}
 \subfloat[]{\includegraphics[scale=1.]{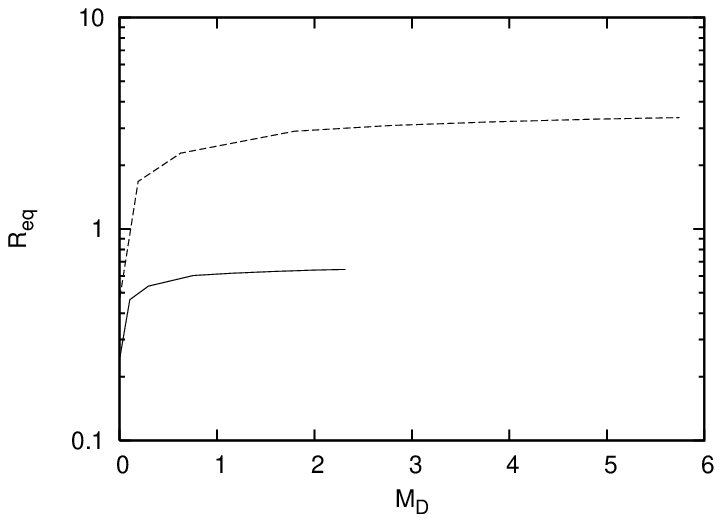}}&
   \subfloat[]{\includegraphics[scale=1.]{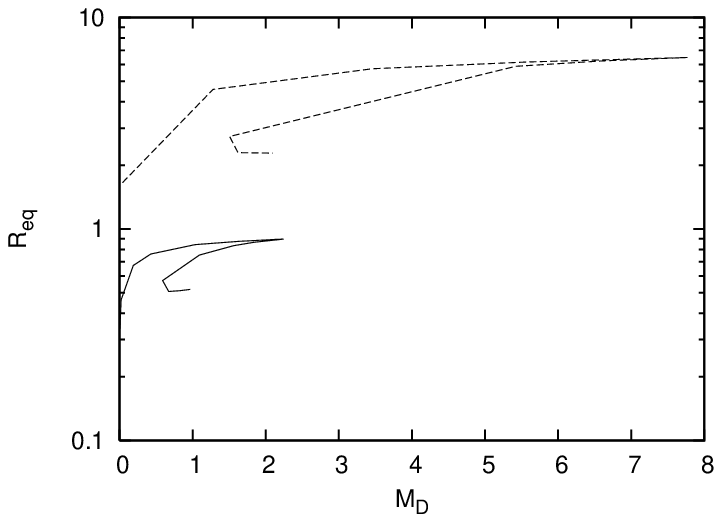}} \\ \\ \\
    \subfloat[]{\includegraphics[scale=1.]{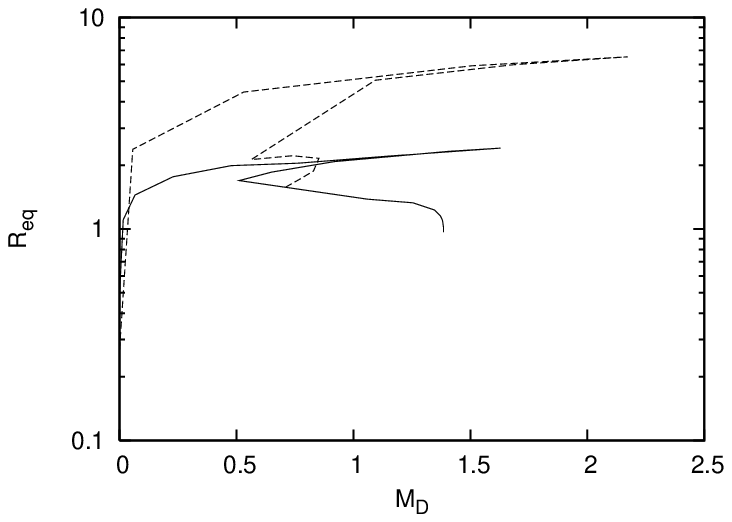}}&
 \subfloat[]{\includegraphics[scale=1.]{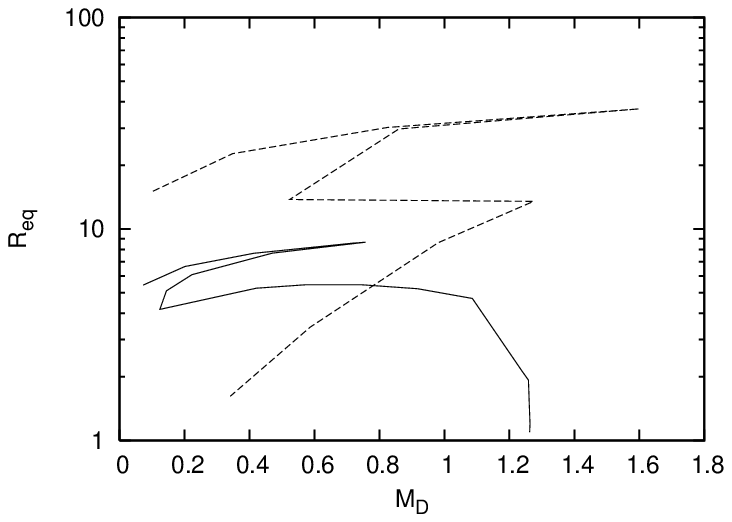}} \\
      \end{tabular}
    \caption{Mass-radius relations with $E_{Fmax} = 20\,m_{e}c^{2}$ for (a) one-level system ($\nu_{m}=1$), (b) two-level system ($\nu_{m}=2$), (c) twenty-level system ($\nu_{m}=20$) and (d) five hundred-level system ($\nu_{m}=500$). Here $M_D$ is the mass of the white dwarf in units of $M_{\odot}$ and $R_{eq}$ is the equatorial radius of the white dwarf in units of $10^{8}$ cm. All the solid lines represent the mass-radius relations for the cases, if the stars would have been spherical. The dashed lines in (a) and (b) represent the mass-radius relations for highly flattened (strongly magnetized) white dwarfs, while the dashed lines in (c) and (d) represent the mass-radius relations for less flattened (relatively weakly magnetized) white dwarfs. In all the four panels the $y$-axis is in log scale.}
    \label{eos}
  \end{center}
\end{figure}

\newpage

\end{document}